\documentclass[aps,epsfig]{revtex4}
\usepackage{graphicx}% Include figure files
\begin{document}
\def\be{\begin{equation}}
\def\ee{\end{equation}}
\def\bfi{\begin{figure}}
\def\efi{\end{figure}}
\def\bea{\begin{eqnarray}}
\def\eea{\end{eqnarray}}

\title{Fluctuation-dissipation relations and field-free algorithms 
for the computation of response functions}

\author{Federico Corberi}
\affiliation {Dipartimento di Matematica ed Informatica and 
INFN, Gruppo Collegato di Salerno, and CNISM, Unit\'a di Salerno,
Universit\`a di Salerno, 
via Ponte don Melillo, 84084 Fisciano (SA), Italy.}
\author{Eugenio Lippiello}
\affiliation {Dipartimento di Scienze Ambientali, Seconda Universit\'a di Napoli,
Via Vivaldi Caserta (Ce), Italy.}
\author{Alessandro Sarracino}
\affiliation {Dipartimento di Matematica ed Informatica, 
Universit\`a di Salerno,
via Ponte don Melillo, 84084 Fisciano (SA), Italy.}
\author{Marco Zannetti}
\affiliation {Dipartimento di Matematica ed Informatica, 
Universit\`a di Salerno, 
via Ponte don Melillo, 84084 Fisciano (SA), Italy.}

\begin{abstract} 

We discuss the relation between the fluctuation-dissipation
relation derived by Chatelain and Ricci-Tersenghi 
[C.Chatelain, J.Phys. A {\bf 36}, 10739 (2003);
F. Ricci-Tersenghi, Phys.Rev.E 68, 065104(R) (2003)] and 
that by Lippiello-Corberi-Zannetti [E. Lippiello, F. Corberi and M. Zannetti
Phys. Rev. E {\bf 71}, 036104 (2005)].
In order to do that, we re-derive the fluctuation-dissipation relation  
for systems of discrete variables evolving in discrete time 
via a stochastic non-equilibrium Markov process. 
The calculation is carried out in a general formalism comprising 
the Chatelain, Ricci-Tersenghi result and 
that by Lippiello-Corberi-Zannetti as special cases.
The applicability, generality, and experimental feasibility of the two approaches is 
thoroughly discussed.
Extending the analytical calculation to the variance of the response function we
show the vantage of field-free numerical methods with respect to the standard method where the perturbation
is applied. We also show that the signal to noise ratio is better
(by a factor $\sqrt 2$) in the algorithm of Lippiello-Corberi-Zannetti
with respect to that of Chatelain-Ricci Tersenghi. 

\end{abstract}

\maketitle


\section{Introduction} \label{intro}

Recently, there has been much interest in the extension in the out of
equilibrium regime of the fluctuation-dissipation theorem (FDT),
through more general fluctuation-dissipation relation (FDR), which have 
led to the concept 
of effective temperature \cite{efftemp} and to the connection between 
non-equilibrium and equilibrium properties \cite{fmpp}.
Fluctuating two-time quantities have also been
actively investigated,
particularly in relation to the detection and quantification
of dynamical heterogeneities, mostly in disordered systems \cite{c4}.

The search of FDR between  response functions  and
properties of the unperturbed system, has led to a number of proposals 
\cite{CKP,Crisanti,chatelain,rt,lcz,Diez,Berthier,noi_nonlin,noi_nonlin2}. 
Among these, the two
by Chatelain \cite{chatelain} and Ricci-Tersenghi \cite{rt} (CRT) and
the one by Lippiello, Corberi, Zannetti \cite{lcz,noi_nonlin,noi_nonlin2} 
(LCZ) have succeeded in making the connection
between the dynamical susceptibility
and unperturbed correlators between observable quantities.
In addition to the intrinsic theoretical interest, these results opened the way to the 
development of perturbation-free numerical algorithms,
allowing for highly efficient and precise measurements of the 
response function via correlators, without need of switching on any perturbation.

However, the paths followed by CRT on one side and by LCZ on the other,
are quite different as well as the final results. The two approaches
lead to expressions of
the susceptibility in terms of radically different unperturbed
correlation functions, making the mapping between them cumbersome.
This poses the question of understanding the 
inner relationship between the two results, of their degree of generality,
of which is performing better in numerical implementations
and of the possible experimental implications.
In this paper we study these issues and answer these questions.
In order to carry out this program, we derive the FDR 
for systems evolving in discrete time via  stochastic Markov processes, 
defined by transition probabilities obeying detailed balance.
We develop a  unified formalism containing 
different approaches as
special cases and explain the difference between those of CRT and LCZ:
while in the LCZ case the response function is related to
correlation functions computed over the whole non-equilibrium ensemble, 
in the CRT approach, instead, averages are taken over a restricted set
of trajectories.

The derivation is fully general for what concerns 
the nature of the discrete variables (e.g. Ising, Potts, Clock etc.) 
and of the transition probabilities. However,
the constraint of a restricted set of trajectories in the CRT approach
requires the microscopic knowledge of the sequence of (attempted) updates, which
is manageable only in numerical simulations.
On the other hand, in the LCZ approach a standard unrestricted ensemble 
average is involved, and the response function is written in terms of
standard correlation functions between observable quantities. This
allows analytical treatments by means of the usual methods 
of statistical mechanics 
and, in principle, experimental applications.
On the other hand, other approaches, such as those in \cite{Crisanti,Diez,Berthier} do not
express the response function in terms of observables.

After clarifying the relations between the FDR in the CRT and LCZ
approaches,
we turn to compare the efficiencies of the numerical
algorithms based on them, together with that of the standard
method (SM), requiring the application of an external perturbation $h$. 
An important advantage of the perturbation-free methods is that the limit
$h\to 0$ is built in, while, with the SM,
checking for linearity is often numerically demanding.
Besides this, field-free methods are also
characterized by a better signal to noise ratio.   
This is a relevant fact, since the numerical computation
of the response function is extremely noisy. 
In order to quantify such a noise, 
we compute exactly the variance
of the response function for each of the three algorithms. 
In the SM it diverges as $1/h^2$, preventing small values of $h$
to be used and making linearity often insecure.
Since, obviously, such a drawback is absent with the field-free methods, 
there is an enormous advantage in their implementation. 
Nevertheless, also with perturbation-free algorithms the noise can be 
significant, especially for large time differences.
The comparison between the variances of the CRT and the LCZ
methods shows that the LCZ approach 
yields a better signal to noise ratio by a factor $\sqrt 2$.
Basically, this  difference is a consequence of the restriction in the set 
of trajectories required by the CRT method.

In addition to the relevance for numerical applications, 
the results for the variances 
give a contribution to the understanding of fluctuations of
two-time quantities,
shedding some light in the field of nonlinear susceptibilities~\cite{bb}. 

After investigating and clarifying the relation between the different algorithms and their performances,
we present the results of numerical simulations in order to discuss
the generality of the method and to illustrate the efficiency
of the field-free algorithms with particular examples.
We compute numerically the response function for models
of Ising spins (the ferromagnetic Ising model and the 
Edwards-Anderson (EA) spin glass in $d=3$) 
with the three methods. We show 
that both the CRT and the LCZ algorithms produce, with great accuracy, the same response 
which can be obtained with the SM. Computing the variances
of the three methods, we obtain the results outlined above.
Finally, we compute the response function in the Fredrickson-Andersen (FA) model,
both applying the perturbation and with the field-free method of LCZ,
finding again perfect agreement. This demonstrates the applicability of the LCZ algorithm
also in this case, and that the criticism raised in Ref. \cite{mayersollich}
does not hold.

This Article is organised as follows:
In Sec.~\ref{model} we present the derivation of the FDR. 
We discuss the results obtained, their generality
and the measurability of the correlators involved. 
In Sec.~\ref{variances} we compute and compare the variances of the three algorithms. 
Sec.~\ref{numerics} is devoted to explicit numerical implementations: We 
consider the $3d$ ferromagnetic Ising and EA models quenched to the critical temperature
and below it. Sec.~\ref{fredand} contains the application to
the FA model. The conclusions are drawn in Sec.~\ref{conclusions}, where some
perspectives are discussed.

\section{Analytical derivation of fluctuation-dissipation relations } \label{model}

We consider a system of $N$ discrete variables $\sigma _i$ (i.e. those 
entering models as Ising, Potts, Clock etc ...), generically
called spins. Time $t$ is discretized, namely $t_n=n\delta$, where $n$ is an integer, and the
time-step is $\delta=1/N$.
A configuration update is attempted at each time step.

\subsection{Transition probabilities}

Spin variables evolve in discrete time according to a generic
Markov chain regulated by the transition probabilities 
$w (\sigma ''\vert \sigma ',n)$ 
to go from a configuration $\sigma '$ to another $\sigma ''$ 
in the $n$-th time-step. Transition probabilities obey the {\it instantaneous} 
detailed balance
\be
w (\sigma ''\vert \sigma ',n)\exp [-\beta {\cal H}(\sigma ',n)]=
w (\sigma '\vert \sigma '',n)\exp [-\beta {\cal H}(\sigma '',n)],
\label{db}
\ee
where ${\cal H}(\sigma,n)$ is the (time dependent) Hamiltonian of the system. 
The diagonal terms
$w (\sigma '\vert \sigma ',n)$ remain fixed by the normalization condition
\be
w (\sigma '\vert \sigma ',n)=1-\sum _{\tilde \sigma \neq \sigma '} 
w(\tilde \sigma \vert \sigma ',n).
\label {norm}
\ee

Restricting, for simplicity, to the case of single spin update,
the form of the transition probabilities at time $n$ is 
\be
w(\sigma ''\vert \sigma ',n)=\frac{1}{N}
\sum _k w_k(\sigma ''\vert \sigma ',n),
\label{dsu}
\ee
where $w_k$ are the single-spin transition probabilities,
namely $\sigma ''$ and $\sigma '$ may differ only for the
$k$-th spin.

The two-time conditional probability 
$P(\sigma,n\vert \sigma',m)$  to go from $\sigma'$ at time $m$
to $\sigma$ at time $n$  
can be expressed as
\be
P(\sigma,n\vert \sigma',m)= \frac{1}{N^{n-m}}\sum_{i_{n-1},\ldots,i_m} 
\sum_{\sigma^{(n-1)},\ldots,\sigma^{(m+1)}}
w_{i_{n-1}}\left (\sigma \vert \sigma^{(n-1)} ,n \right )\ldots 
w_{i_{m}}\left (\sigma^{(m+1)} \vert \sigma' ,m \right ).
\label{pnm}
\ee
In the case of time independent $w$, the conditional probability is
time translation invariant. For later use, we write this property 
as 
\be
P(\sigma,n\vert \sigma',m+1)=P(\sigma,n-1\vert \sigma',m).
\label{tti}
\ee

Given two generic observables $A(\sigma)$ and $B(\sigma)$ (namely functions of
a configuration of the system), 
from the knowledge of the conditional probability one can compute
their correlation function
\be
C_{AB}(n,m)=\langle A(n)B(m)\rangle=
\sum_{\sigma,\sigma'}A(\sigma)
P(\sigma,n\vert \sigma',m)B(\sigma ')P(\sigma',m).
\label{observ}
\ee

\subsection{Relation between perturbed and unperturbed transition
probabilities}

In the presence of an external perturbation $h_j(n)$ switched-on in
the $j$-th site, the evolution is controlled by the Hamiltonian 
${\cal H}(\sigma ,n )={\cal H}_0(\sigma)-\sigma _jh_j(n)$.
In the following we will always consider time-independent unperturbed
transition probabilities and we will drop the time dependence in the
unperturbed transition rates.  
The detailed balance condition~(\ref{db})
for the perturbed transition probabilities reads

\be
\frac{w_j^h(\sigma''|\sigma',n)}{w_j^h(\sigma'|\sigma'',n)}=
\frac{e^{-\beta {\cal H}_0(\sigma'')}}{e^{-\beta {\cal H}_0(\sigma')}}e^{\beta h_j(n)(\sigma_j''-\sigma_j')},
\label{dbh1}
\ee
where, from now on, $w_j$ and $w_j^h$ refer to unperturbed and 
perturbed transition probabilities, respectively.
The most general form of $w^h_j$ obeying Eq.~(\ref{dbh1}) is

\be
w_j^h(\sigma''|\sigma',n)=w_j(\sigma''|\sigma',n)e^{\frac{\beta}{2}h_j(n)(\sigma''_j-\sigma'_j)}
M_j(\sigma',\sigma'',n),
\label{dbh2}
\ee
where $M_j(\sigma',\sigma'',n)$ is an $h$-dependent function symmetric with 
respect to the exchange of its arguments and such that
$w_{j}^h$ is a probability, namely positive and normalizable.
To linear order in the external perturbation one has
\begin{eqnarray}
w_{j}^h(\sigma''\vert \sigma',n)&=&w_{j}(\sigma''\vert \sigma')\left [
1-\frac {\beta }{2}h_{j}(n)(\sigma_{j}'-\sigma_{j}'')+m_{j}(\sigma'',\sigma')h_j(n)\right ] 
(1-\delta _{\sigma',\sigma''})  \nonumber \\
&+&\left \{ 1-\sum _{\tilde \sigma \neq \sigma'}
w_{j}(\tilde \sigma \vert \sigma',n)\left [
1-\frac {\beta }{2}h_{j}(n)(\sigma_{j}'-\tilde \sigma _{j})+
m_{j}(\tilde \sigma ,\sigma',n)h_j(n)\right ] \right \}
\delta _{\sigma',\sigma''}
\label{transh}   
\end{eqnarray}
where 
\be
m_{j}(\sigma'',\sigma',n)= \left .
\frac{\partial M_j(\sigma'',\sigma',n)}{\partial h_j(n)}
\right \vert _{h=0} .
\label{mpiccolo}
\ee 
Let us comment on $m_j$: It is well known that the detailed balance condition 
leaves an arbitrariety on the choice of the transition probabilities, both
for the unperturbed and the perturbed ones. Even for a fixed
choice of unperturbed $w_j$, therefore,
there is a family of different $w_j^h$ obeying detailed balance, parametrized by $m_j$.

\subsection{Response function} \label{response}

For a magnetic perturbing field $h_j(m)$ turned on the site $j$ 
in the $m$-th time-step 
the impulsive response function $R_{i,j}(n,m)$, describing the effect
of the perturbation on the spin $\sigma_i$ at time $n>m$, is defined by
\be
R_{i,j}(n,m)=
\frac{1}{\delta }\left . \frac {\partial
  \langle \sigma _i(n)\rangle}{\partial h_j(m)} \right| _{h=0} =N\left . 
\frac {\partial \langle \sigma _i(n)\rangle}{\partial h_j(m)}
\right| _{h=0} ,
\ee   
where averages $\langle \ldots \rangle$ are taken over thermal histories and the initial
condition. 

From Eq.(\ref{pnm}), one has
\be
R_{i,j}(n,m)=
N \sum_{\sigma,\sigma',\sigma''}\sigma_i
P(\sigma,n\vert \sigma'',m+1)
\left . \frac{d w^h_{j}\left (\sigma'' \vert \sigma' \right )}
{d h_j } \right \vert _{h=0} P(\sigma',m).
\label{rij2}
\ee
The derivative of the $w_k^h$
with respect to the field can be easily obtained from Eq.(\ref{transh})
\be
\left . \frac{\partial w_j^h(\sigma ''\vert \sigma ',n)}{\partial h_j(m)}\right |_{h=0}=
w_j(\sigma ''\vert \sigma ')\left [f_j(\sigma '',\sigma ')
(1-\delta _{\sigma ',\sigma ''})
+g_j(\sigma')
\delta _{\sigma ',\sigma ''} \right ]\delta_{n,m} 
\label{dpdh}   
\ee
where 

\be
f_j(\sigma '',\sigma ')=
-\frac {\beta }{2}(\sigma '_j-\sigma ''_j)+
m_j(\sigma '',\sigma '),
\label{f}   
\ee 
and 
\be
g_j\left (\sigma \right ) 
w_{j}
\left (\sigma \vert \sigma \right )  =
-\sum_{\sigma' \ne \sigma}w_{j}\left (\sigma' \vert \sigma \right ) 
f_{j}\left (\sigma' , \sigma \right ). 
\label{gj}
\ee

From Eqs. (\ref{dpdh},\ref{f},\ref{gj}) it is clear that the response function
of Eq. (\ref{rij2}) cannot be straightforwardly interpreted
as correlation functions. In order to do that, one would need
the full transition probability 
$P(\sigma,n\vert \sigma'',m+1)w(\sigma''|\sigma',m)$ connecting
$\sigma'$ at time $m$ to $\sigma''$ at time $n$, with $w(\sigma''|\sigma',m)$ 
containing all the 
$w_k(\sigma''|\sigma')$ according to Eq.(\ref{dsu}), 
while in Eqs.~(\ref{dpdh}) only the \emph{one site} $w_j(\sigma''|\sigma')$ appears.
A way out is to insert the missing $w(\sigma''|\sigma',m)$ by writing
$Ndw^h_j/dh_j|_{h=0}=dw^h/dh_j|_{h=0}=wd (\ln w^h)/dh_j|_{h=0}$, 
as proposed in \cite{Berthier}, obtaining
\be
R_{i,j}(n,m)=
\sum_{\sigma,\sigma',\sigma ''}\sigma_i
P(\sigma,n\vert \sigma'',m+1)w(\sigma '' |\sigma ')
\left . \frac{d \ln w ^h\left (\sigma'' \vert \sigma' \right )}
{d h_j } \right \vert _{h=0} P(\sigma',m).
\label{berth}
\ee
However, let us notice that, although the response function is expressed
in terms of the unperturbed 
dynamics, the function appearing on the r.h.s. of Eq. (\ref{berth})
is not in the form of a correlation function
between observables according to
the definition (\ref{observ}). This is because $d (\ln w^h)/dh_j|_{h=0}$
depends on two configurations.
 
Going back to Eq. (\ref{rij2}), in order to illustrate the CRT and the LCZ 
approaches, it is useful to write the response function as 
the sum of an off-diagonal contribution $ \overline D_{i,j}(n,m)$  
and a diagonal contribution $D_{i,j}(n,m)$ 
\be
R_{i,j}(n,m)= \overline D_{i,j}(n,m)+D_{i,j}(n,m) 
\label{Rdd}
\ee
with 
\be
\overline  D_{i,j}(n,m)= 
N \sum_{\sigma,\sigma',\sigma ''} \sigma_i
P(\sigma,n\vert \sigma'',m+1) 
w_{j}\left (\sigma'' \vert \sigma' \right ) 
f_{j}\left (\sigma'' , \sigma' \right )
\left [1-\delta_{\sigma'',\sigma'}\right ]  P(\sigma',m)
\label{d1}
\ee
and
\be
 D_{i,j}(n,m)= 
N \sum_{\sigma,\sigma',\sigma ''}
\sigma_i P(\sigma,n\vert \sigma'',m+1) 
w_{j}\left (\sigma'' \vert \sigma' \right ) 
g_{j}\left (\sigma' \right )
\delta_{\sigma'',\sigma'}  P(\sigma',m).
\label{d2}
\ee

The above equations are exact and fully general. 
The next step is to express $D_{i,j}$ and $\overline D_{i,j}$
in terms of correlation functions of observable quantities. 
This can be done in two different ways, 
leading to the CRT and LCZ results. We describe them
separately below.  

\subsection{CRT class algorithms}

Given the time interval $(n,m)$, in numerical simulations one fixes
a sequence ${\cal I}(n,m)$ of sites to be updated and then sums over
different sequences. 
This corresponds to rewrite the conditional probability~(\ref{pnm}) in the form 
\be
P(\sigma,n\vert \sigma',m)=\frac{1}{N^{n-m}}\sum_{{\cal I}(n,m)} 
\sum_{\sigma^{(n-1)},\ldots,\sigma^{(m+1)}}
w_{I(n-1)}\left (\sigma \vert \sigma^{(n-1)} \right )\ldots
w_{I(m)}\left (\sigma^{(m+1)} \vert \sigma' \right ),
\label{pnm2}
\ee   
where the sum extends over all $N^{n-m}$ possible choices of ${\cal
  I}(n,m)$ in the interval $[m,n]$.  
Hence, $(1/N)P(\sigma,n\vert \sigma'',m+1)
w_{j}\left (\sigma'' \vert \sigma' \right )$ is the
conditional probability restricted on the ensemble of trajectories
satisfying the constraint $I(m)=j$, where $I(m)$ is the particular
site updated at time $m$ in a given trajectory. 
This implies that
$\overline D_{i,j}(n,m)$ can be written as the correlation 
$\langle\sigma_i(n)f_j(m)\delta_{I(m),j}\rangle_{flip}$
between $\sigma_i$   and $f_j$,
taking into account only trajectories where the $j$-th spin has been
flipped at time $m$. Similarly, $D_{i,j}(n,m)$ is the correlation 
$\langle\sigma_i(n)g_j(m)\delta_{I(m),j}\rangle_{no flip}$
between $\sigma_i$ and $g_j$ including only trajectories where flipping
$\sigma_j$ has been attempted at time $m$ but rejected.
Hence, the response function can be written as

\be
R_{i,j}(n,m)= N\langle\sigma_i(n)f_j(m)\delta_{I(m),j}\rangle_{flip} + 
N\langle\sigma_i(n)g_j(m)\delta_{I(m),j}\rangle_{no flip}.
\label{flipnoflip}
\ee
This result is fully general. It holds irrespective of the nature of the discrete variables and of the
form of the transition probabilities $w$ and $w^h$.
Notice that, because of the $\delta$ function, on average only one out of $N$ trajectories
contributes to $R_{i,j}$. Therefore the overall factor $N$ makes $R_{i,j}$ well defined
in the $N\to \infty $ limit.

Chatelain \cite{chatelain} and Ricci-Tersenghi \cite{rt} have considered the
particular case of Ising spins 
interacting via the Hamiltonian $H(\sigma,n)=-\sum _i \sigma _i [H^W_i(\sigma)+h_i(n)]$,
where $H_j^W (\sigma)= J\sum _{<i>_j}\sigma _i$ is the Weiss field (the sum runs
over the spins 
interacting with $\sigma_j$), and of
heat-bath transition probabilities  
\be
w_j^h(\sigma ' \vert \sigma,m )=
\frac {\exp [\beta (H_j^W(\sigma)+h_j(m))\sigma _j']}
{2\cosh [\beta (H_j^W(\sigma)+h_j(m))]}.
\label{hb}
\ee
This specific choice corresponds to
\be
f_{j}\left (\sigma, \sigma' \right ) =\beta(\sigma_j-\sigma_j^W)=
g_{j}\left (\sigma \right ),
\label{fjgj}
\ee
where $\sigma_j^W=\tanh(\beta H_j^W)$,
allowing to rewrite $R_{i,j}(n,m)$ in the more compact form
\be
R_{i,j}(n,m)= N\beta \langle \sigma_i(n)
\left [\sigma_j(m+1)-\sigma_j^W(m)\right ] \delta_{I(m),j} \rangle.
\label{ricci}
\ee
Here, since $f_j=g_j$, the distinction between $\langle\ldots\rangle_{flip}$
and $\langle\ldots\rangle_{no flip}$ in Eq.~(\ref{flipnoflip}) can be avoided.
Although $R_{i,j}$ in Eq.(\ref{ricci}) is related to averages in the
unperturbed dynamics, the $\delta_{I(m),j}$ acts like a projector on 
the restricted ensemble of
phase-space trajectories including an attempted update 
of $\sigma _{j}$ at time $m$. 
This is also the ensemble of trajectories that contributes to
$R_{i,j}(n,m)$ in standard numerical simulations where the
perturbation is applied. 
The presence of the projector $\delta _{I(m),j}$ makes necessary  
the knowledge of the sequences of updated spins 
restricting the applicability
of this FDR to numerical simulations.
This problem is bypassed in the LCZ algorithm, as shown below.

\subsection{LCZ algorithm}

In this section we re-derive the results of refs. \cite{lcz,noi_nonlin2},
originally obtained in a continuous time formalism,
in the case of evolution in discrete time.
Starting from the definition~(\ref{d2}) of $D_{i,j}$
and using the time translation invariance property~(\ref{tti}),
one has $D_{i,j}(n,m)=\sum_{\sigma,\sigma'}
\sigma_i P(\sigma,n-1 \vert \sigma',m) 
B_j(\sigma') P(\sigma',m)$ 
where 
$B_j(\sigma)=(\beta /2)w_j(\sigma \vert \sigma) g_j(\sigma)$.
Hence,
\be
D_{i,j}(n,m)=\langle \sigma_i(n-1)B_j(m) \rangle.
\label{ddij4}
\ee
We stress that, differently from the CRT scheme of Eq. (\ref{ricci}),
the above form implies that no projection over a restricted ensemble
of trajectories is present.

We now turn to consider $\overline D_{i,j}$.
To begin with, taking advantage of the arbitrariness of $m_j$,
let us consider the simplest choice $m_j=0$ in 
Eq. (\ref{transh}). The effects of different choices of
$m_j$ will be considered in
Sec. \ref{mdiff0}. Then, from Eq.~(\ref{f}) 
one has $f_j(\sigma'',\sigma')=-(\beta /2)(\sigma_j'-\sigma_j'')$ and,
since $\sigma'$ and $\sigma''$ may differ at most for the spin on site $j$,
one can write 
\be
\frac{1}{N}w_{j}\left (\sigma'' \vert \sigma' \right )
\left (\sigma_j''-\sigma_j' \right )=
\frac{1}{N}\sum_k w_{k}\left (\sigma'' \vert \sigma' \right )
\left (\sigma_j''-\sigma_j' \right )= 
w\left (\sigma'' \vert \sigma' \right )
\left (\sigma_j''-\sigma_j' \right ),
\label{wtot}
\ee 
showing that $w_j$ can be replaced with the full transition probability.
Inserting into Eq.~(\ref{d1}), $\overline D_{i,j}(n,m)$ takes the form 
\be
\overline D_{i,j}(n,m)= 
\frac{\beta}{2} \langle \sigma_i(n)
\Delta\sigma_j(m) \rangle,
\label{overd}
\ee
where
\be
\Delta \sigma_j(m)=N[\sigma_j(m+1)-\sigma_j(m)]
\label{deltassigma}
\ee
allows one to identify the discrete time derivative with respect to $m$ 
of the autocorrelation function $C(n,m) = 
\langle \sigma_j(n)\sigma_j(m) \rangle$ in Eq. (\ref{overd}). 

In conclusion, with the choice $m_j=0$ made in \cite{lcz},
one has the relation
\be
R^{LCZ}_{i,j}(n,m)=\frac{\beta}{2} \left [ \langle
\sigma_i(n)\Delta \sigma_j(m)\rangle  -
\langle \sigma_i(n-1) B_j(m) \rangle \right ],
\label{LCZf}
\ee   
with
\be
B_i(\sigma)=\sum_{\sigma'} w_i(\sigma' \vert \sigma)(\sigma'_i-\sigma_i),
\label{bi}
\ee
which is the form usually considered in the applications \cite{lcz,lavorinumerici,nat}.
Notice that $B$ depends on a single configuration and hence the
term involving it in Eq. (\ref{LCZf}) is a correlation between observable
quantities. 

As stressed previously, the above result, in addition to being general 
with respect to the form of the
single spin flip unperturbed 
transition probabilities, holds true
\cite{lcz} also for transition     
probabilities involving multiple-spin updates (as, for instance, Kawasaki
spin-exchange). Extensions to the response of generic observables 
and to the case of
transition probabilities that do not
obey detailed balance are discussed in \cite{noi_nonlin,noi_nonlin2} 
and in \cite{nat}, respectively.

In Eq. (\ref{LCZf}), at variance with the CRT result, no reference is made
to the site $I(m)$ to be updated at time $m$, and therefore there is
no 
restriction on the ensemble of trajectories to be considered.
The average over
all possible choices of $I(m)$ is, therefore, analytically performed.
As it will be shown in Sec. \ref{variances} 
this makes the LCZ  more efficient in numerical applications.
More important, being an ordinary non equilibrium average,
Eq. (\ref{LCZf}) is well suited to standard analytical calculations and,
in principle, to experiments.

Finally, let us point out a property of correlations involving $B_i$ 
that will be useful in the following.
Given a generic observable $O(m)$ at a time $m\le n-1$, 
from the definition~(\ref{bi}) one has
\be
\langle B_i(n) O(m) \rangle  = \langle 
\Delta \sigma_i(n) O(m) \rangle .
\label{b2}
\ee 
Indeed,
\be
\langle B_i(n) O(m) \rangle = 
\sum_{\sigma, \sigma', \tilde \sigma} w_i(\tilde \sigma \vert
\sigma)(\tilde \sigma_i-\sigma_i)P(\sigma,n\vert \sigma',m)O(\sigma')
P(\sigma',m)
\ee
and using Eq.~(\ref{wtot}) one obtains Eq.~(\ref{b2}).
Eq. (\ref{b2}) shows that in the mean $B_i$ plays the role of the time derivative of a spin.

\subsection{Extra contributions related to $m_j \ne 0$ }\label{mdiff0}

We now explore the consequences of a different choice
of $m_j\neq 0$ within the the LCZ
scheme. Retaining the $m_j$ contributions in Eq.~(\ref{transh}),
the response function can be written as 
\be
R_{i,j}(n,m)=R_{i,j}^{LCZ}(n,m)+\epsilon_{i,j}(n,m)
\label{duer}
\ee
with 
\bea
& & \epsilon_{i,j}(n,m)=
\beta \sum_{\sigma,\sigma'',\sigma'} 
\sigma_iP(\sigma,n\vert \sigma'',m+1) 
\Big \{ 
w_{j}(\sigma''    \vert \sigma') 
m_{j}(\sigma'',\sigma') 
\left [1-\delta_{\sigma'',\sigma'}\right ] \nonumber \\ 
& &
+ \sum_{\tilde \sigma\ne \sigma'}
m_j(\sigma'',\sigma')
w_{j}\left (\tilde \sigma \vert \sigma' \right ) \delta_{\sigma'',\sigma'}
\Big \}P(\sigma',m).
\label{epsi} 
\eea
Since the full transition probability cannot be reconstructed as in Eq.~(\ref{wtot}),
$\epsilon _{i,j}$ can be identified as a correlation only in the restricted
phase space of trajectories with the $j$-th spin updated at the time
$m$ as in the CRT scheme. The choice $m_j=0$, therefore, has the advantage of avoiding this problem.

It must be stressed that the formal manipulations leading to Eqs.~(\ref{ricci})
and~(\ref{duer})
are exact and hence they are
identical if the same transition probabilities $w^h$, or equivalently the same
choice of $M$, are considered. In particular, Eq. (\ref{duer})
contains the CRT relation of refs. \cite{chatelain,rt} as a particular case
when heat-bath transition probability~(\ref{hb}), corresponding to $m_j(\sigma'',\sigma')=-\sigma_j^W(\sigma')$
is used.

Therefore, let us compare the CRT and LCZ results in this case.
Observing that, from Eq.(\ref{norm}),
$\sum _{\tilde \sigma\ne \sigma'}w_{j}\left (\tilde \sigma \vert \sigma'
\right )=1-w_j(\sigma'\vert \sigma')$ and using Eq.~(\ref{tti}), one has
\be
\epsilon_{i,j}(n,m)=-N\langle \sigma_i(n) \sigma_j^W(m) \delta_{I(m),j}
\rangle + \langle \sigma_i(n-1) \sigma_j^W(m) \rangle  .
\label{epsi2}
\ee
The first observation is that $\epsilon_{i,j}(n,m)=0$ in
equilibrium. Indeed, from the definition of $\sigma ^W_j$ one has
$\langle \sigma _i(n)
\sigma_j^W(m) \delta_{I(m),j}\rangle =\langle \sigma _i(n)
\sigma_j^W(m+1) \delta_{I(m),j}\rangle$. Then, we use time reversal invariance
to exchange the time arguments. The $\delta $ function acting now at the
larger time can be replaced by a factor $1/N$ representing the fraction of 
contributing trajectories. Finally, exchanging again the time arguments, 
one obtains 
$N\langle \sigma_i(n) \sigma_j^W(m) \delta_{I(m),j}
\rangle=\langle \sigma_i(n-1) \sigma_j^W(m) 
\rangle$ 
and the right hand side of the above equation vanishes.
Out of equilibrium this is no more
true. However, it is generally expected that large-scale, long-time properties
of scaling systems in the thermodynamic limit are not affected
by the precise form of transition probabilities, provided detailed
balance hold. The effect of different choices of $m_j$, therefore, is expected
to be negligible. Numerical simulations, presented in the next Section, 
confirm the expectation.

%Nevertheless, because of local equilibrium,
%$\epsilon_{i,j}(n,m)\simeq 0$ except at interfaces (questo vale
%sicuramente nel coarsening magari si uo' fare piu' in generale). This
%implies that $\epsilon_{i,j}(n,m)$ is roughly proportional to interface
%density and then is a decreasing function of $m$. In the response
%function, conversely,  the bulk response is non null and this makes 
%$\epsilon_{i,j}(n,m)$ negligible with respect to $R_{i,j}(n,m)$.

\section{Variances}
\label{variances}

The FDR's of CRT and LCZ have opened the way to numerical algorithms
for the computation of the response function without applying the
perturbation, the so called field-free methods. 
It was shown in Refs.~\cite{lavorinumerici,rt}
that the calculation of the response function made via the CRT and LCZ algorithms 
is very precise and numerically efficient.

In this Section
we compute analytically the variances of the fluctuations of the
response function obtained with the standard method (SM) where the perturbation
is switched on, and with the two field-free methods. This
task is carried out for Ising spins, the CRT method being valid only in this case.
This allows us to compare the 
numerical efficiency of the different algorithms and to comment on
the physical relevance of the variances, particularly in the context of
systems with quenched disorder(see Section~\ref{integrated}).

Let us start by defining the fluctuating response function $r_{i,j}$ by
\be
R_{i,j}(n,m)=\langle r_{i,j}(n,m)\rangle
\ee
and, therefore, its variance by 
\be
\Delta^{(R)}_{i,j}(n,m)=R^{(2)}_{i,j}(n,m)-R^2_{i,j}(n,m)
\label{varvar}
\ee
where
\be
R^{(2)}_{i,j}(n,m)=\langle r_{i,j}(n,m) r_{i,j}(n,m) \rangle.
\label{r2}
\ee

We then focus on $R^{(2)}_{i,j}$, computing it separately in the three methods.

\begin{enumerate}

\item{
{\it Standard method}

In the standard method one applies a sufficiently small magnetic field $h$
at time $m$ in the $j$-th site, 
and the response function is obtained by numerically implementing
Eq. (\ref{rij2}) where
\be
\left .\frac{d w_j^h(\sigma''\vert \sigma')}{dh_j} \right \vert_{h=0}=
\frac{w_j^h(\sigma''\vert \sigma')-w_j(\sigma''\vert \sigma')}{h_j}.
\label{dwh}
\ee
Since $w_j^h(\sigma''\vert \sigma')$ enters Eq. (\ref{rij2}) 
as the probability to flip $\sigma _j$ at the time $m$,
the above numerical derivative takes contribution different from zero
only on the ensemble of trajectories were at that time the update of $\sigma_j$ is
attempted. Then, imposing this restriction by means of
the projector $\delta_{I(m),j}$ and taking into account that $\langle
\sigma_i(n) \rangle=0$ in the unperturbed dynamics, one obtains 
\be
R_{i,j}(n,m)=N\frac{\langle \sigma_i(n)\delta_{I(m),j}\rangle_h}{h}
\label{rsm}
\ee
where the average is over the perturbed dynamics.
We next observe that 
$(\delta_{I(m),j})^2=\delta_{I(m),j}$ and that from Eq. (\ref{pnm2})
$\langle \delta_{I(m),j} \rangle=1/N$, since the $\delta$ function
cancels the sum over $I(m)$ in $P(\sigma,n\vert \sigma',n)$. From
Eq. (\ref{rsm})  one then  obtains   
\be
R^{(2)}_{i,j}(n,m)=\frac{N}{h^2} . 
\label{smfluc}
\ee
Notice that $R^{(2)}_{i,j}$
diverges in the $h \to 0$ limit.}

\item{
{\it CRT relation}

Form Eq.~(\ref{flipnoflip}) one obtains
\be
R^{(2)}_{i,j}(n,m)= N^2\beta^2\left \langle 
f_j ^2\left (\sigma(m+1) , \sigma(m)\right )   \delta _{I(m),j} \right \rangle_{flip}+
N^2\beta^2\left \langle g_j^2\left(\sigma(m)\right ) 
\delta _{I(m),j} \right \rangle _{no flip},
\label{rijCRT7}
\ee
which holds true for any choice of $m_j$. Notice that the ensembles of
trajectories contributing to the averages $\langle \ldots \rangle_{flip}$ and
$\langle \ldots \rangle_{no flip}$ are orthogonal. Therefore, no cross terms
are present in Eq.~(\ref{rijCRT7}). 
Restricting to the case of Ising spins and heat-bath transition probability, using
Eq.~(\ref{fjgj}) one finds
\be
R^{(2,CRT)}_{i,j}(n,m)=\beta^2 N^2 
\langle \left [ 1-2\sigma_j(m+1) \sigma_j^W(m)+\sigma_j^W(m)^2 \right
] \delta _{I(m),j} \rangle
. \label{r2crt}
\ee
}

\item{

{\it LCZ relation}

In this case  $r_{i,j}$ is given in Eq.~(\ref{LCZf}), and one has
\be
R^{(2,LCZ)}_{i,j}(n,m)= 
\langle r_{i,j}(n,m)r_{i,j}(n,m) \rangle= 
\frac{\beta^2}{4}  
\langle \Delta \sigma_j^2(m) \rangle+
\frac{\beta^2}{4}  
\langle  B_j^2(m)\rangle-
\frac{\beta^2}{2}  
\langle \Delta \sigma_j(m) B_j(m)\rangle .
\label{r22}
\ee
Notice that $\langle \Delta \sigma_j^2(m)\rangle=2N^2\langle 1-\sigma_j(m+1)
\sigma_j(m)\rangle=2N^2\langle w_j(\sigma'\neq \sigma | \sigma)\delta _{I(m),j}\rangle$, 
where the last
equality holds because only trajectories where at time $m$ the $j$-th spin is flipped
give a non-vanishing contribution. Hence, since the $\delta $ function 
contributes on average only once every $N$ trajectories, $\langle \Delta \sigma _j ^2 (m)\rangle \propto N$.  
The second term on  the r.h.s. of Eq. (\ref{r22}) does not depend on $N$.
Regarding the third term, reasoning along the same lines as for  $\langle \Delta \sigma_j^2(m)\rangle$,
from the definition (\ref{deltassigma}) it follows that it is independent on $N$. 
Then, neglecting the last two terms in the large-$N$ limit one has
\be
R^{(2,LCZ)}_{i,j}(n,m)=  
\frac{\beta^2}{4}  
\langle \Delta \sigma_j^2(m) \rangle=
\frac{\beta^2}{2} N^2  
\langle 1-\sigma_j(m+1) \sigma_j(m) \rangle
\label{r2LCZ}
\ee
}

\end{enumerate}

\subsection{Comparison among variances}

As already mentioned, in the limit $h \rightarrow 0$ the standard method leads to
a diverging variance.

We now compare the variances of the two
field-free methods using in both cases heat-bath unperturbed
transition probabilities~(\ref{hb}).   
We first observe  that (see Appendix I)
\be
1-\langle \sigma_j(m+1) \sigma_j(m) \rangle=
\frac{1}{N} \left [1-
\langle \sigma_j(m)\sigma_j^W(m) \rangle \right ]
\label{ssss}
\ee
and therefore
\be
R^{(2,LCZ)}_{i,j}(n,m)=  
\frac{\beta^2}{2} N  
\langle 1-\sigma_j^W(m) \sigma_j(m) \rangle .
\label{r2LCZ2}
\ee

Next, for heat-bath transition probability, it can be shown
(see Appendix I) that
\be
\langle \sigma_j(m+1)\sigma_j^W(m) \delta_{I(m),j} \rangle=
\frac{1}{N}\langle \sigma_j^W(m)^2 \rangle .
\label{sssw}
\ee
Using the above result in Eq.~(\ref{r2crt}), we get
\be
R^{(2,CRT)}_{i,j}(n,m)=
\beta^2 N  
\langle 1-\sigma_j^W(m)^2 \rangle ,
\label{r2CRT2}
\ee
and finally, comparing with Eq.(\ref{r2LCZ2}),  
one obtains
\be
R^{(2,CRT)}_{i,j}(n,m)=2 R^{(2,LCZ)}_{i,j}(n,m)
+N\beta ^2\left \langle \sigma_i^W(m) \left [\sigma_i(m)-\sigma_i^W(m) \right ]
\right \rangle.
\label{r2CRT2f}
\ee

Recalling Eq.~(\ref{epsi2}), and using Eq. (\ref{sssw}), one can show that 
$N\beta ^2\left \langle \sigma_i^W(m) \left [\sigma_i(m)-\sigma_i^W(m) \right ]
\right \rangle=\epsilon _{j,j}(m+1,m)$. As discussed in Sec. \ref{mdiff0},
this term is zero in equilibrium and one expects it to be negligible also out of equilibrium
(this fact will be checked by numerical simulations in Sec. \ref{numerics}).
Then one has 
\be
R^{(2,CRT)}_{i,j}(n,m)\simeq2 R^{(2,LCZ)}_{i,j}(n,m).
\label{r2CRT2fb}
\ee
In order to compute the variances, according to Eq. (\ref{varvar})
the term $R^2_{i,j}$ should be subtracted from $R^{(2)}_{i,j}$.  
However, these terms are negligible with respect to
$R^{(2)}_{i,j}$ in the thermodynamic limit being independent on $N$.  
Hence
\be
\Delta^{(R,CRT)}_{i,j}(n,m)\simeq 2\Delta^{(R,LCZ)}_{i,j}(n,m).
\label{varvera2}
\ee

The numerical evaluation of fluctuation of response functions
confirms the above result, as it will discussed in Section \ref{numerics}. 

The origin of the factor 2 in the variances can be related to the
different ways the term $D_{i,j}$ of Eq. (\ref{d2}) is treated in the CRT and LCZ
methods and, in particular, to the presence of the $\delta$-function in
the CRT scheme (\ref{ricci}).
Indeed, from Eq. (\ref{flipnoflip}) one has that, in the CRT scheme,
the fluctuating part of both $D_{i,j}$ and $\overline D_{i,j}$
(corresponding to the two terms on the r.h.s) are
non vanishing only once every $N$ trajectories, and in this case their
contribution is of order $N$. Therefore, both the contributions to the variance 
coming from $D_{i,j}$ and  $\overline D_{i,j}$ are
of order $N$.
Conversely, in the LCZ scheme, this is true only for the
term $\overline D_{i,j}$ of Eq. (\ref{overd}), because 
$\Delta \sigma _j$ in Eq. (\ref{deltassigma}) is of order $N$ only
once every $N$ trajectories when the $j$-th spin is flipped. 
Instead, from Eq. (\ref{ddij4})
one has that all trajectories provide a term of order one to $D_{i,j}$.
So, the contribution to the variance associated with this term
is of order one, and hence negligible.
In conclusion, in the thermodynamic limit there are two terms contributing 
in the same way to the variance for the CRT algorithm whereas only one survives 
in that of LCZ.

\subsection{Integrated response function}
\label{integrated}

As already mentioned, the measurement of the impulsive response function $R$ 
is numerically very demanding, so, in order to reduce the noise, usually the
time integrated response function (dynamic susceptibility) is considered
\be
\chi _{i,j}(n,m)=\frac{1}{N}\sum _{l=m}^nR_{i,j}(n,l)=\langle x_{i,j}(n,m)\rangle,
\label{intrespf} 
\ee
where $x_{i,j}$ is the fluctuating part of $\chi _{i,j}$.  
In numerical simulations we focus on the equal site integrated response
$\chi_{i,i}$.  Taking advantage of space translation invariance, one usually computes   
the spatial average $x (n,m)\equiv (1/N)\sum _{i=1}^N x _{i,i}(n,m)$ which fluctuates less
than $x_{i,i}$.
The variance of this quantity can be written
in the form
\be
\Delta ^{(\chi)}(n,m)=\Delta ^{(\chi)}_{0}(n,m)+ \Delta ^{(\chi)}_{r}(n,m),
\ee
where 
\be
\Delta ^{(\chi)} _0(n,m)=\frac{1}{N^2} \sum _{i=1}^N 
\langle x_{i,i}^2 \rangle 
-\frac{1}{N}\chi^2(n,m)
\label{delta0}
\ee
contains only equal sites terms,
and 
\be
\Delta ^{(\chi)} _r(n,m)=\frac{1}{N^2}\sum _{i\neq j}
\langle x_{i,i} x_{j,j} \rangle 
-\frac{N-1}{N}\chi^2(n,m)
\label{deltar}
\ee
is the contribution from different sites.

In the case of simulations with the external field, the standard
procedure consists in switching on a random perturbation during the
interval $[n,m]$. One generally uses the  bimodal distribution
$\overline {h_ih_j}=h^2\delta_{i,j}$ where the over-line
indicates averages over the external perturbation. The integrated
response function is then given by \cite{prev2}
\be
\chi(n,m)=\frac{1}{N h^2} \sum_{i=1}^N \overline{\langle \sigma_i
  \rangle_h h_i}
\label{eq57}
\ee
where $\langle \rangle_h$ is the average in the presence of the
perturbation.
From the above equation and using the bimodal distribution of the
external field one   obtains
\be
\Delta ^{(\chi)} _0(n,m)=\frac{1}{N h^2} 
-\frac{1}{N}\chi^2(n,m)
\label{delta0SM}
\ee
 and
\be
\Delta ^{(\chi)} _r(n,m)=\frac{1}{N^2 h^4}\sum _{i\neq j}
\overline {\langle s_{i} s_{j} \rangle _h h_i h_j} 
-\frac{N-1}{N}\chi^2(n,m),
\label{deltarSM}
\ee
that can be also written as
\be
\Delta ^{(\chi)}_r(n,m)=\sum_{i\ne j}\chi^{(2,2)}_{i,j}(n,m),
\label{deltarchi22}
\ee
where 
\be
\chi^{(2,2)}_{i,j}(n,m)=\frac{1}{N^2}\sum_{l=m}^n\sum_{l'=m}^n
\left .\frac{\partial^2 \langle\sigma_i(n)\sigma_j(n)\rangle}{\partial
h_i(l)\partial h_j(l')}\right |_{h=0}-\chi_{i,i}(n,m)\chi_{j,j}(n,m),
\label{chi22}
\ee
is a second order susceptibility. This quantity  
represents a  tool for identifying  cooperative effects
in disordered systems, as it was proposed in  ~\cite{huse,noi_nonlin,noi_nonlin2}
and checked numerically in ~\cite{noi_nonlin,noi_nonlin2}.  

We then turn to consider the algorithms without the probing
field. We first observe that the term $\Delta_r$ is identical for all
the algorithms and is always related to the non-linear susceptibility
$\chi^{2,2}_{i,j}$, via Eq.(\ref{deltarchi22}). Indeed, from the
definition (\ref{deltar}) one has
\be
\Delta ^{(\chi)} _r(n,m)=\frac{1}{N^4}\sum _{i\neq j}
\sum _{l=m}^n\sum _{l'=m}^n
\langle r_{i,i}(n,l)r_{j,j}(n,l')\rangle
-\frac{N-1}{N}\chi^2(n,m) .
\label{deltarr}
\ee
The term $\langle  r_{i,i}(n,l)r_{j,j}(n,l)\rangle$ in Eq. (\ref{deltarr}) 
can be written, using Eq. (\ref{rij2}), as
\be
\langle  r_{i,i}(n,l)r_{j,j}(n,l)\rangle
=\sum_{\sigma,\sigma'',\sigma',\tilde\sigma,\tilde\sigma'}\sigma_i\sigma_j
P(\sigma,n\vert \sigma'',l+1)
\left . \frac{d w^h_{i}(\sigma'' \vert \sigma')}
{d h_i }\right|_{h=0}
P(\sigma',l\vert \tilde\sigma,l'+1)\left . \frac{d w^h_{j}(\tilde\sigma \vert \tilde\sigma')}
{d h_j }\right | _{h=0} P(\tilde\sigma',l').
\ee
The r.h.s. of this equation can be readily interpreted as the second order response
$R^{(2,2)}_{i,j}(n,l,l')=\left .\frac{\partial^2 \langle\sigma_i(n)\sigma_j(n)\rangle}{\partial
h_i(l)\partial h_j(l')}\right|_{h=0}$ leading to
Eq.(\ref{deltarchi22}).
On the other hand, the term $\Delta_0$ has  different behaviors for
the different algorithms. As already shown, $\Delta_0$ diverges as $h
\to 0$ in the standard method. We now explicitly  consider the term
$\Delta_0$ in the CRT and LCZ algorithm. From the definition   
\be
\Delta ^{(\chi)}_0(n,m)=\frac{1}{N^4}\sum_{i=1}^N\sum _{l=m}^n\sum _{l'=m}^n
\langle r_{i,i}(n,l)r_{i,i}(n,l')\rangle
-\frac{1}{N}\chi^2 (n,m).
\ee
As shown in Appendix II, $\langle r_{i,j}(n,l)r_{i,j}(n,l')\rangle=0$ 
for $l\neq l'$ and, therefore,  
only the terms with $l'=l$   contribute in the double sum in this equation, yielding 
\be
\Delta ^{(\chi)} _0(n,m)=\frac{1}{N^4} \sum _{i=1}^N \sum_{l=m}^n
R^{(2)}_{i,i}(n,l)
-\frac{1}{N}\chi^2(n,m).
\label{delta00}
\ee
In both the CRT and LCZ 
algorithms $\Delta ^{(\chi)}_0$ is an increasing function of time
roughly proportional to $n-m$, as already pointed out in \cite{rt}.
This follows from substituting Eq.~(\ref{r2LCZ})
into Eq.~(\ref{delta0}) and using Eq.~(\ref{sssw}), obtaining
\be
\Delta ^{(\chi)}
_0(n,m)=\frac{\beta^2}{N}\sum_{l=m}^n[1-\langle\sigma_i^W(l)^2 \rangle]-
\frac{1}{N}\chi^2(n,m).
\label{deltachi000}
\ee
Since, at finite temperature, $\langle\sigma_i^W(l)^2\rangle$ is strictly
less than one, the first term gives a contribution growing as $n-m$ whereas
$\chi^2$ is always at most equal to $\beta^2$, 
and then sub-dominant at large times. Therefore,
from the result~(\ref{varvera2}) at large times one has
\be
\Delta ^{(\chi,CRT)}_0(n,m)= 2\Delta^{(\chi,LCZ)}_0(n,m).
\label{delta02}
\ee
The numerical analysis supporting this result is presented in the following Section.

\section{Numerics} \label{numerics} 

In this Section we present the results of the numerical computation of the
integrated response function $\chi _{i,i}$ of Eq. (\ref{intrespf}) using the SM with
heat bath transition probabilities, and the CRT and 
LCZ methods. Details on the numerical implementation of the algorithms
are given in Appendix III.
 
On the basis of the analysis of Sec. \ref{model}, since the LCZ algorithm
corresponds to a different choice of $M_j$ in Eq. (\ref{dbh2}) with respect
to the other two, one may expect some differences in the results.
However the numerical data presented below show that the response function computed
with all the methods is the same within the numerical uncertainty.  
We then compute the variances of the response function, in order to check the analysis
discussed in Sec. \ref{variances} and to quantify the performance of the different methods. 
In the second part of this Section
we also present data for the Fredrickson-Andersen model, showing that
the LCZ field-free algorithm
can be successfully applied also in this case.   

\subsection{Ising and EA models} \label{isingmodel} 

We consider $N=100^3$ Ising spins interacting via the Hamiltonian 
${\cal H}(\sigma)=-\sum_{\langle i,j\rangle}J_{ij}\sigma_i\sigma_j$,
where the sum runs over nearest-neighbour spins on a three-dimensional cubic lattice
with periodic boundary conditions,
evolving according to heath-bath transition probabilities.
The quantity $B$ entering Eq.~(\ref{LCZf}) takes the form
\be
B_i(\sigma)=\sigma^W_i-\sigma_i.
\label{eq68}
\ee
In particular we focus on the ferromagnetic Ising model ($J_{ij}=J=1$) and on the
EA model ($J_{ij}=\pm 1$ with equal probability). Temperature is measured in units of $J$. 

\begin{figure}
    \centering
    %\vbox to 8.5 cm {
   \rotatebox{0}{\resizebox{.43\textwidth}{!}{\includegraphics{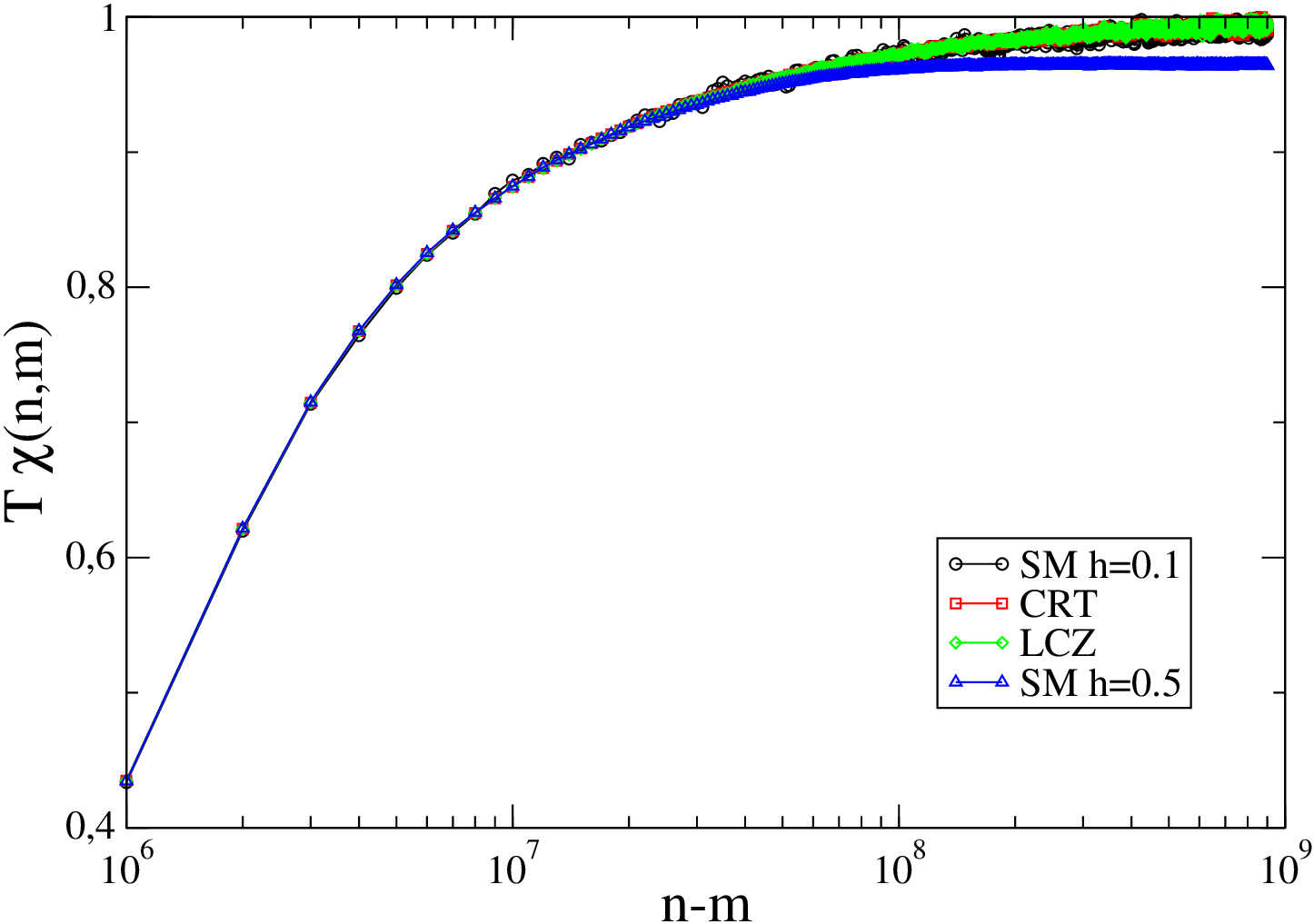}}} 
\rotatebox{0}{\resizebox{.45\textwidth}{!}{\includegraphics{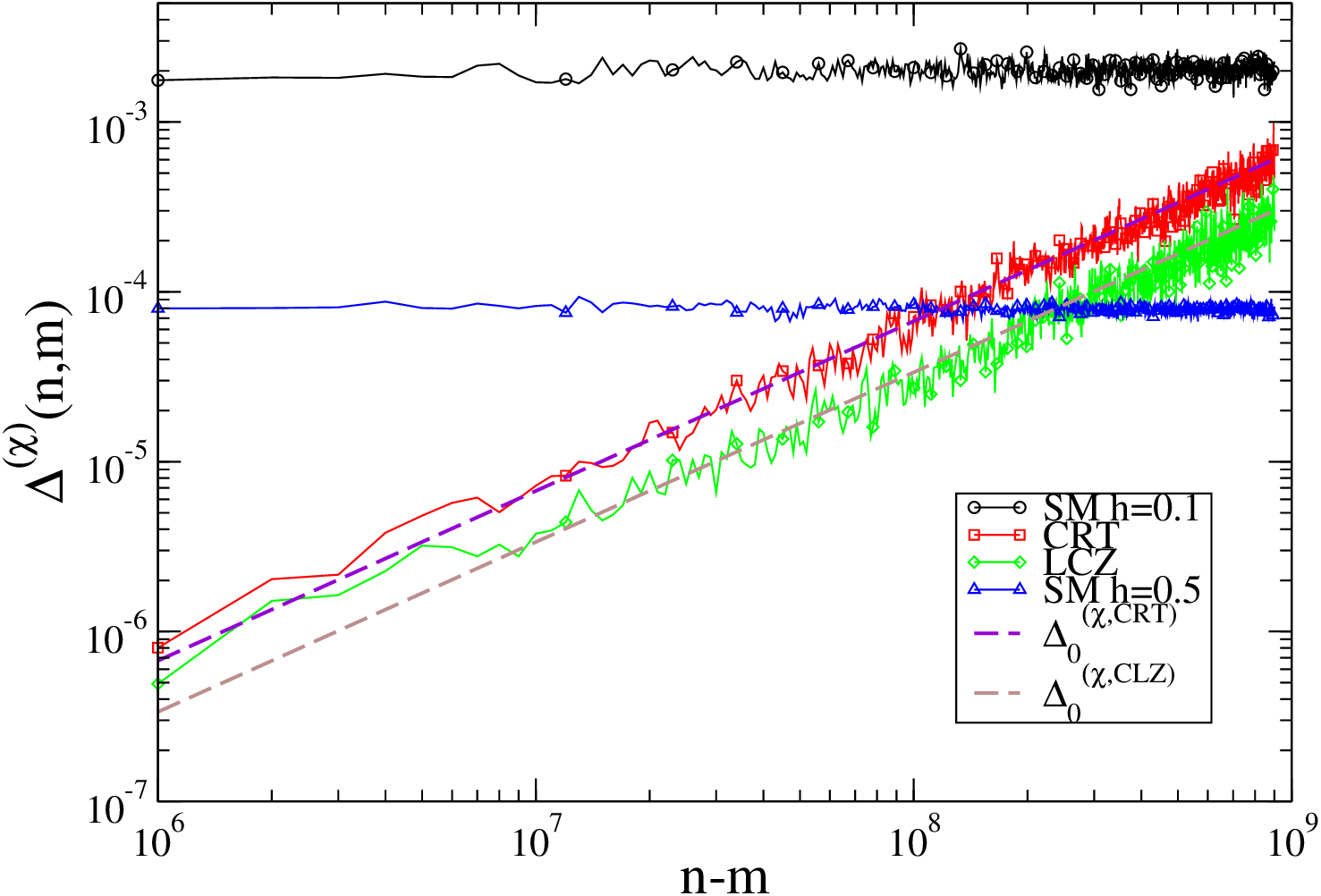}}}
    \caption{ (Color online). Integrated auto-response function 
    $\chi (n,m)$ (left panel) and its variance (right panel)
    in the 3-dimensional Ising model quenched from $T=\infty$ to $T_c=4.5115$.
    Left panel: Different curves correspond to computations performed with the SM (using
    two different values of $h$) and with the field-free algorithms of CRT and
    LCZ, as indicated in the key.
    In the right panel the behavior of the variances of the
    response function computed with the different methods is shown (continuous lines
    with heavy symbols, see key). The terms $\Delta ^{(\chi,CRT)}_0$
    and $\Delta ^{(\chi,LCZ)}_0$ are also plotted (dashed lines).}
\label{ising}
\vspace{1cm}
\end{figure}

Let us start with the Ising model. In Fig.~\ref{ising}
the off equilibrium evolution of the system after a quench from infinite
temperature to $T_c=4.5115$ is considered. The susceptibility 
computed with the CRT and LCZ methods, and with the 
SM with $h=0.1$ and $h=0.5$ is plotted against $n-m$ in  the left panel. 
As it can be seen, the first three computations yield
the same result with good accuracy. The SM with $h=0.5$, instead, 
agrees with the other cases only up to $n-m \simeq 3 \cdot 10^7$.
 This
shows that non-linear effects become important from $n-m\simeq 3 \cdot 10^7$ onwards. 
$\chi (n,m)$ grows monotonously to
the equilibrium value $T\chi _{eq}=1$ as already found in previous studies~\cite{noiteff}.

The right panel shows the variances $\Delta ^{(\chi)}$. Results with the SM, both with $h=0.1$ and $h=0.5$, 
give a time-independent value very well consistent with
$\Delta ^{(\chi,SM)}=\Delta ^{(\chi,SM)}_{0}=1/(Nh^2)$ (see Eq.~\ref{delta0SM}). 
This implies that $\Delta ^{(\chi,SM)}_{r}$ is
negligible and then $\Delta ^{(\chi,SM)}_{0}$ dominates the fluctuation
of $\chi$.
With the CRT and LCZ algorithms one finds variances which are
proportional to each other, within the numerical uncertainty, i.e. 
$\Delta ^{(\chi,CRT)}(n,m)\simeq 2\Delta ^{(\chi,LCZ)}(n,m)$
in agreement with Eq.~(\ref{varvera2}),
and, as expected,
they grow approximately linearly in $n-m$.  
According to the analysis of Sec. \ref{integrated}
this implies that $\Delta ^{(\chi)} _{r}$ is negligible with respect to 
$\Delta ^{(\chi)} _{0}$.
Actually, this can be checked in Fig.~\ref{ising},
where the term $\Delta ^{(\chi)} _{0}$ alone is plotted, showing that it substantially coincides
with the whole variance $\Delta ^{(\chi)} $ both for CRT and LCZ.

Notice that, since $\Delta ^{(\chi,SM)} $ is constant while $\Delta ^{(\chi,CRT)}$ and
$\Delta ^{(\chi,LCZ)}$ grow in time, $\Delta ^{(\chi,SM)}$ becomes
smaller than the other two for large times, as it can be seen in the case $h=0.5$ in Figure~\ref{ising}.
However, when this happens, the data obtained with the SM are already 
affected by large non-linear effects, as it is evident from the difference between 
the curves corresponding to $h=0.1$ and $h=0.5$, in the left panel. Hence, the 
signal to noise ratio in the field-free methods is better than the one in 
the SM, provided that one works in the linear regime, as it is shown in the right panel. 
Moreover, the factor $2$ between the variances implies that,
in order to have a certain signal to noise ratio,
simulations performed with the CRT algorithm require a 
larger statistics (by a factor $\sqrt 2$) than those based on
the LCZ method.

\begin{figure}
    \centering
    %\vbox to 8.5 cm {
   \rotatebox{0}{\resizebox{.43\textwidth}{!}{\includegraphics{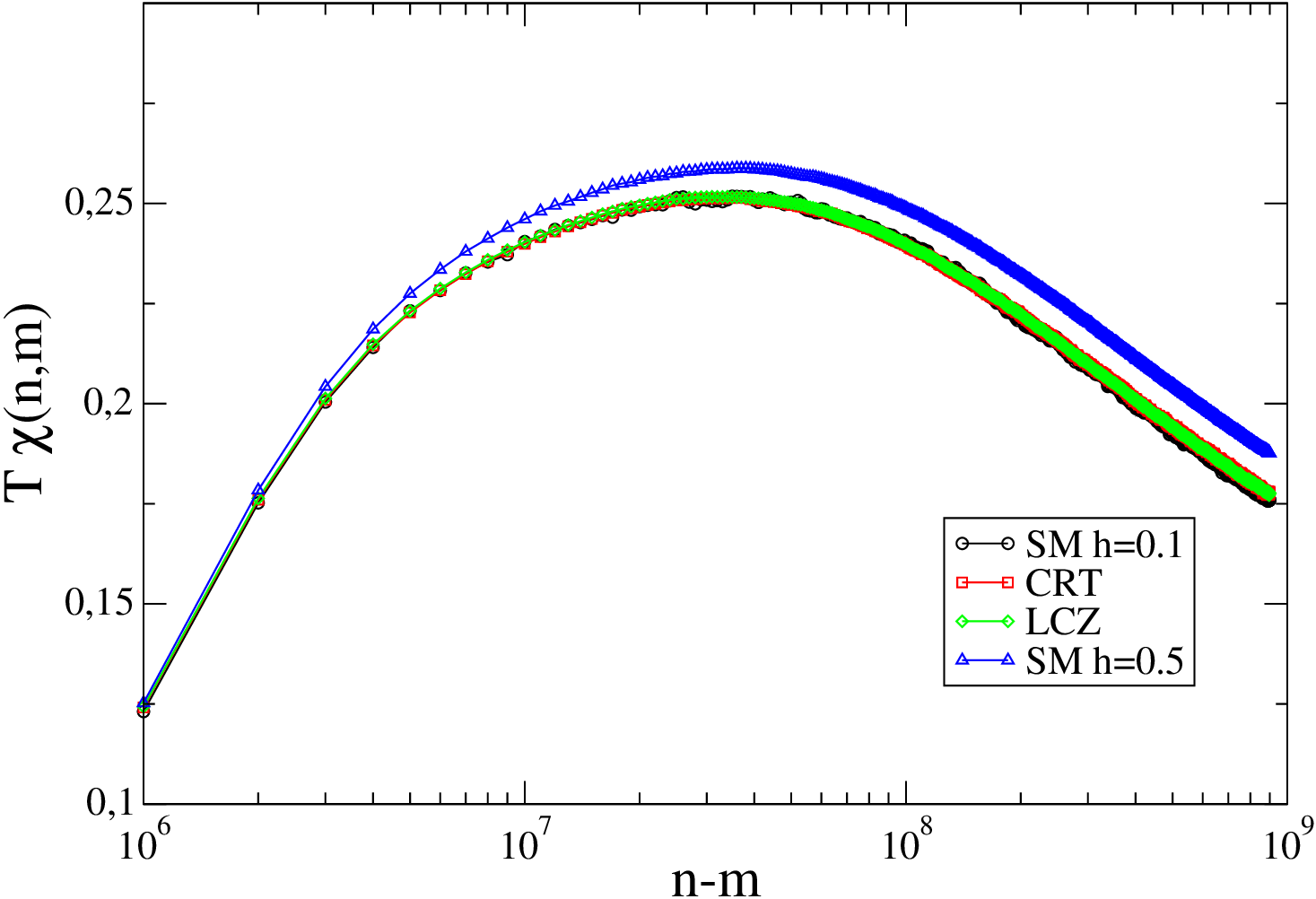}}}
   \rotatebox{0}{\resizebox{.45\textwidth}{!}{\includegraphics{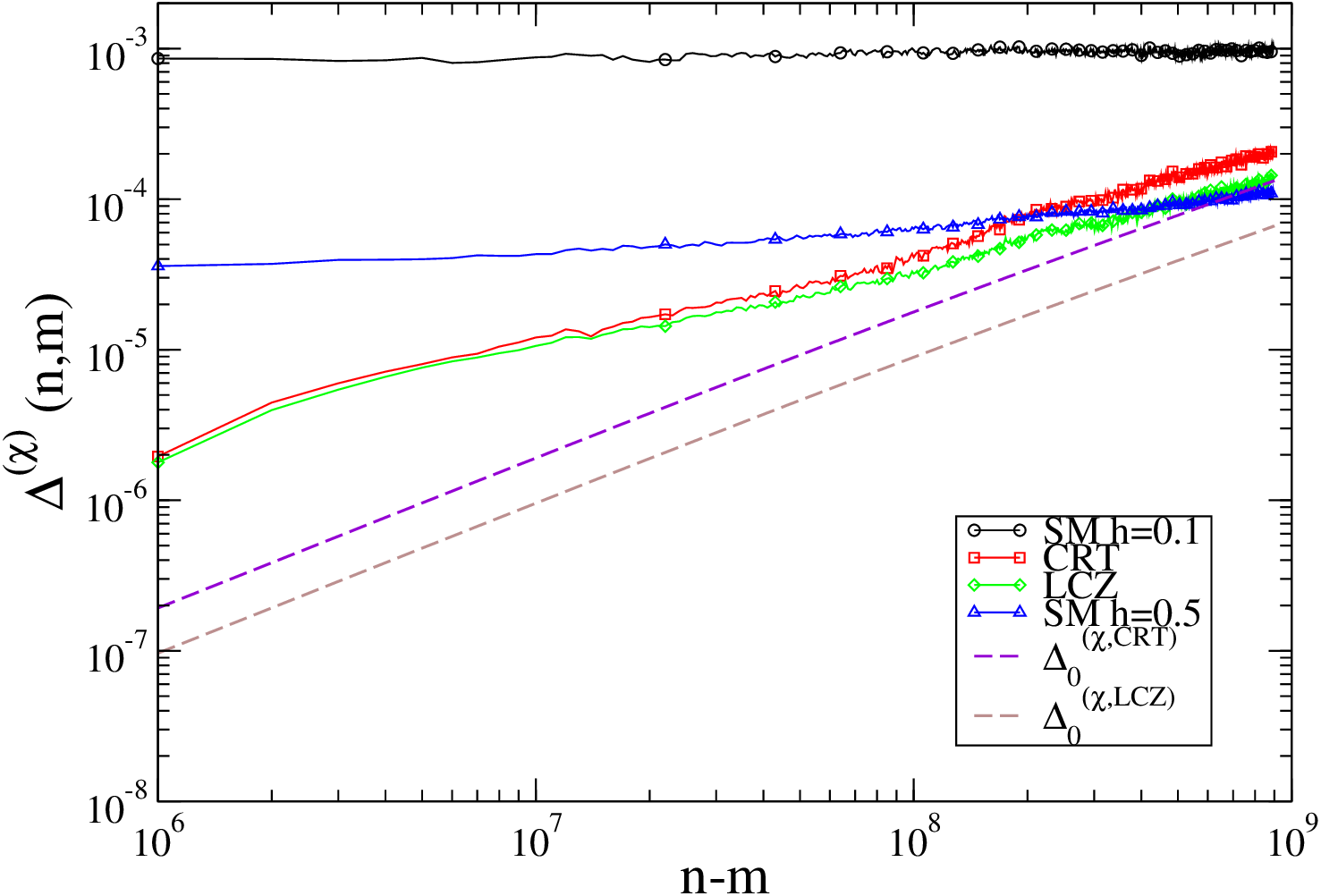}}}
    \caption{(Color online). As in Figure \ref{ising} for  
the 3-dimensional Ising model quenched from
    infinite temperature to $T=3<T_c$.
    Left Panel:  Different algorithms  yield the same result 
    except the SM with the largest value of $h$ which, is always 
    affected by non-linear effects. }
\label{ising2}
\vspace{1cm}
\end{figure}

\begin{figure}
    \centering
    %\vbox to 8.5 cm {
   \rotatebox{0}{\resizebox{.45\textwidth}{!}{\includegraphics{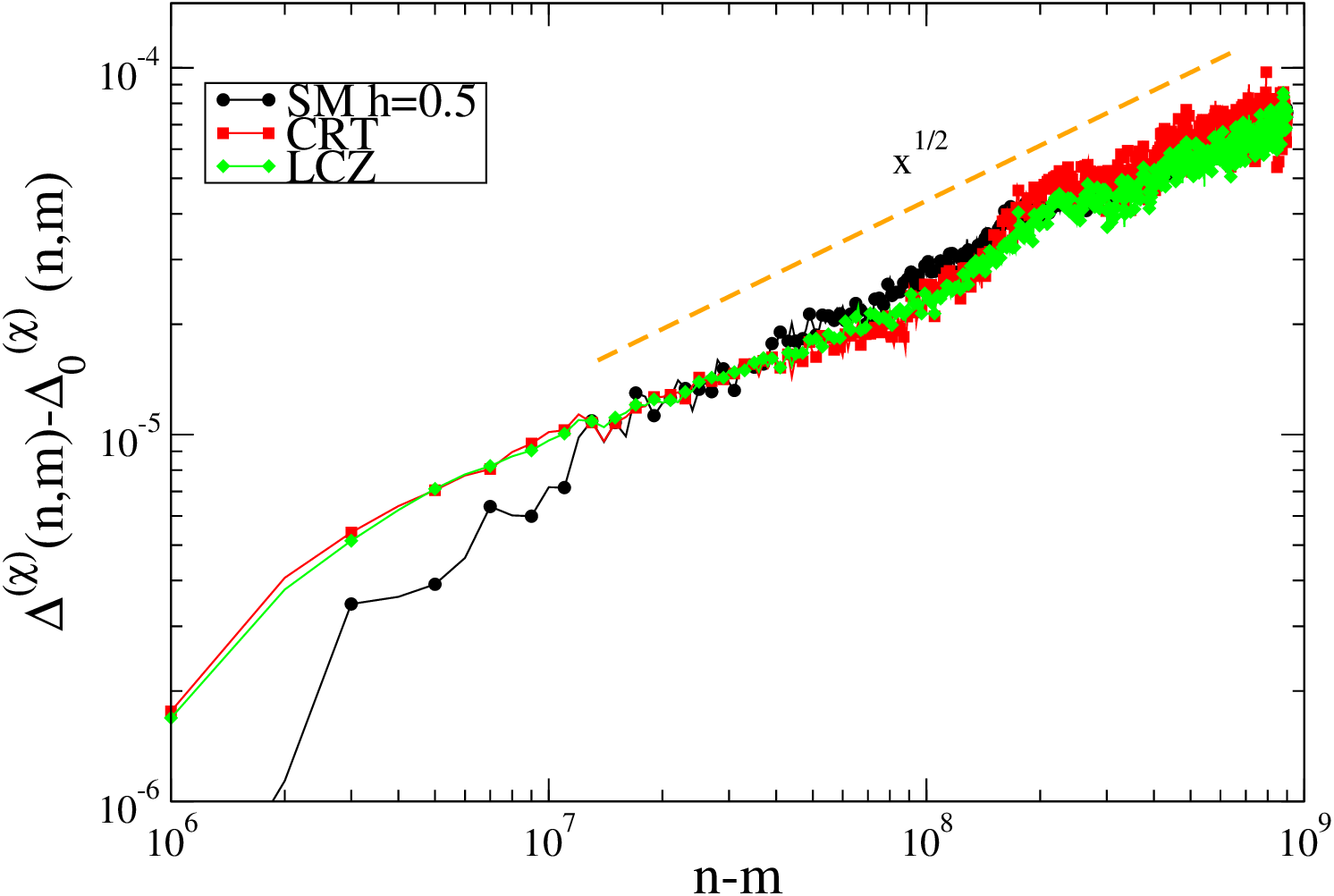}}}
    \caption{(Color online). Difference $\Delta ^{(\chi)}(n,m)-\Delta ^{(\chi)}_0(n,m)$ for the three algorithms
      in the Ising model quenched to $T=3$. The dashed orange line has slope 0.5.} 
\label{figdelta}
\end{figure}

Fig.~\ref{ising2} displays the behavior of the same quantities as in
Fig.~\ref{ising} in the case of a quench to $T=3$, below $T_c$.
The behavior of the susceptibility is now characterized by a maximum
around $n-m\simeq m$, due to the interplay between the 
response of single interfaces and the reduction of their number~\cite{prev2},
as domain coarsening goes on. 
For what concerns the comparison between the variances, we first observe that in the SM the variance is almost
time-independent around a value in good agreement with $1/(Nh^2)$. 
With the field-free methods, $\Delta ^{(\chi)}_0(n,m)$ still grows linearly with $n-m$ and
the relation~(\ref{delta02}) is very well verified.
Nevertheless, at variance with the quench to $T_c$,
$\Delta ^{(\chi)}_0$ does not represent the whole variance $\Delta ^{(\chi)}$  
and the contribution $\Delta ^{(\chi)} _r$ is not negligible.
As already mentioned, we expect  $\Delta ^{(\chi)} _r$ to be the same for all 
the methods. This is shown in Fig.~\ref{figdelta}, where $\Delta ^{(\chi)} _r=\Delta ^{(\chi)}-\Delta ^{(\chi)}_0$
is plotted against $n-m$. The behavior of $\Delta ^{(\chi)} _r$ is 
consistent with the power law $\Delta ^{(\chi)} _r(n,m)\propto (n-m)^{0.5}$.  
Therefore, since the growth of $\Delta ^{(\chi)} _r$ is slower than that observed for
$\Delta ^{(\chi)} _0$, the proportionality
$\Delta ^{(\chi,CRT)}\simeq 2\Delta^{(\chi,LCZ)} $ between the whole variances is expected to
hold at times larger than those accessed in the simulation.

The observed difference in the behavior of $\Delta ^{\chi}$ in the
quench to $T_c$ and to below $T_c$ can be attributed to different
structure of domains. In the quench to below $T_c$, domains are
compact and inside domains $\langle \sigma_i^W(m)^2 \rangle \simeq
m_{eq}^2$ where $m_{eq}$ is the equilibrium value of the magnetization
in the ordered phase. Therefore, indicating with $\rho(t)$ the defect
density and taking into account that at interfaces $\langle
\sigma_i^W(m)^2 \rangle \simeq 0$
one roughly expects $\langle \sigma_i^W(m)^2 \rangle \simeq
m_{eq}^2(1-\rho(t))$. Substituting this result into
Eq. (\ref{deltachi000}) 
one obtains asymptotically, when $\rho(t) \ll 1$,
 \be
\Delta^{(\chi)}_0(n,m) \simeq \left (1-m_{eq}^2\right ) (n-m).
\label{delta0linear}
\ee

Since in the
ordered phase $m_{eq}$ is very close to one, this implies that 
$\Delta^{(\chi)}_0$ linearly grows in time, as indeed observed, but with a very small
prefactor (the smaller the lower is the temperature).  
This makes $\Delta^{(\chi)}_0$ comparable or even
sub-dominant at small time differences $n-m$ with respect to 
$\Delta^{(\chi)}_r$. Nevertheless, since $\Delta^{(\chi)}_r$ grows with a smaller exponent
($\simeq 0.5$)  fluctuations are always dominated by $\Delta^{(\chi)}_0$ at
large times. Conversely, in the case of quenches to $T=T_c$,
$m_{eq}=0$ and $\Delta^{(\chi)}_0$ is the dominant contribution even for small
$n-m$.

\begin{figure}
    \centering
    %\vbox to 8.5 cm {
       \hspace{1cm}
   \rotatebox{0}{\resizebox{.45\textwidth}{!}{\includegraphics{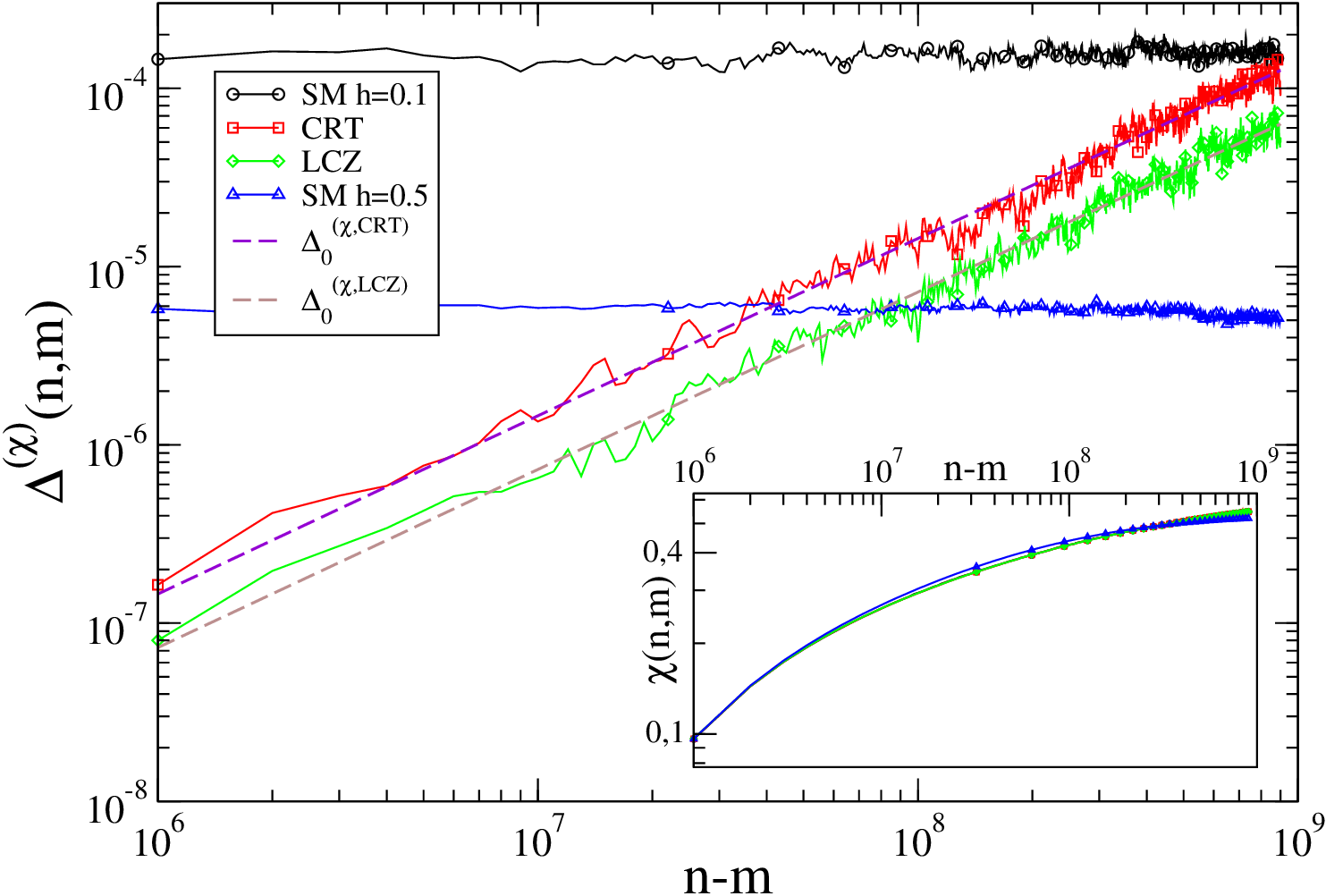}}}
   \rotatebox{0}{\resizebox{.45\textwidth}{!}{\includegraphics{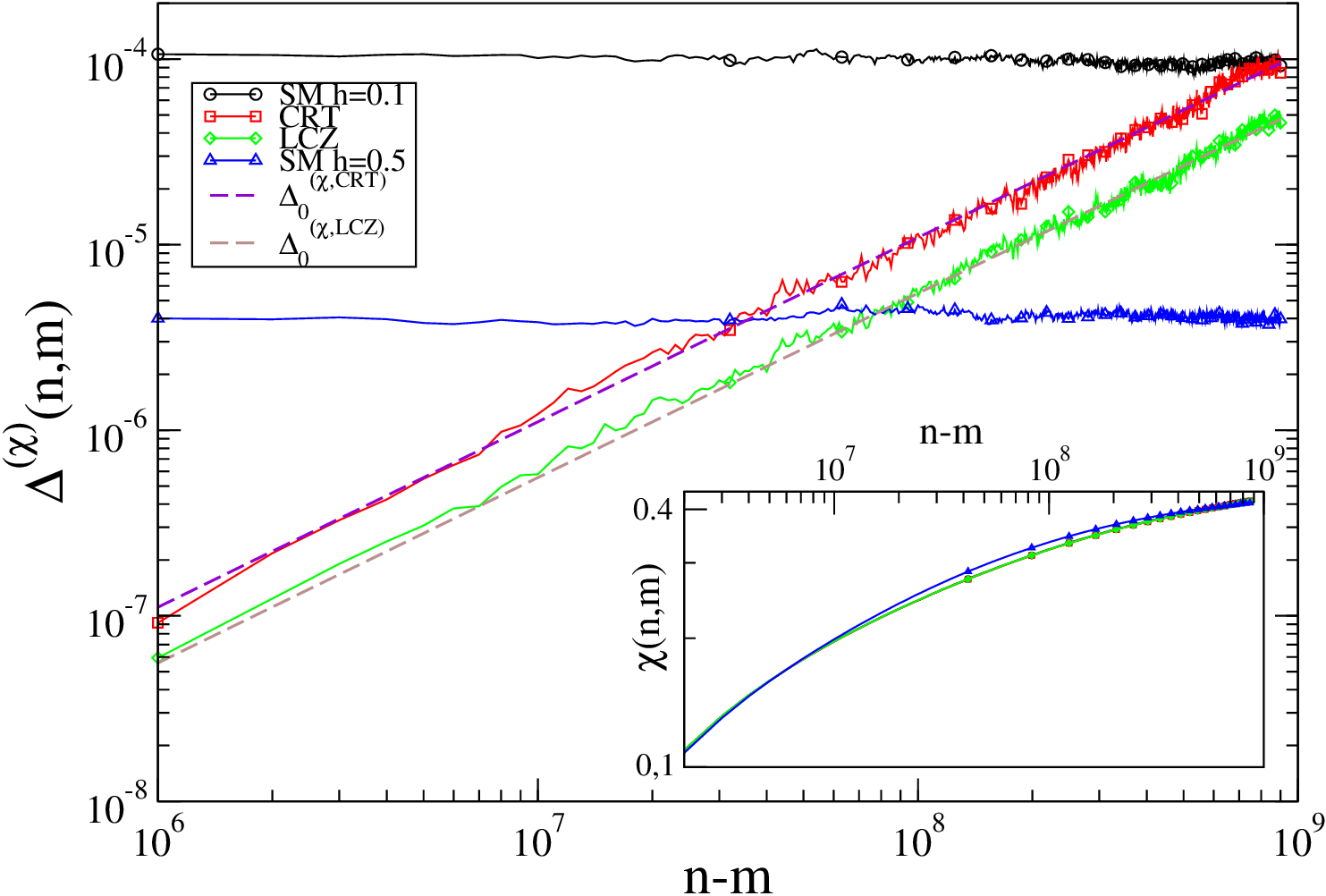}}}
   \vspace{1cm}
    \caption{(Color online).
    The integrated auto-response function
    $\chi (n,m)$ (insets) and its variance (main)
    in the 3-dimensional EA model quenched from
    $T=\infty $ to $T_c\simeq 1.2$ (left panel) or to $T=1<T_c$.
    Insets: Different curves correspond to computations performed with the SM and
    two different values of $h$ and with the field-free algorithms of CRT and
    LCZ, as indicated in the key. 
    In the main part of the two panels the behavior of the variances of the
    response computed with the different methods is shown (continuous lines
    with heavy symbols, see key). The terms $\Delta ^{(\chi,CRT)}_0$
    and $\Delta ^{(\chi,LCZ)}_0$ are also plotted (dashed lines).}
\label{ea}
\vspace{1cm}
\end{figure}

In Fig.~\ref{ea} we show the integrated response function and its variance 
in the EA model quenched to $T_c$ (left) and below $T_c$ (right).
Here we take for $T_c$ the value obtained in ref.~(\cite{eatc}).
All the methods yield the same result 
except the SM with the largest value of $h$ which, for large values of $n-m $ 
($n-m \gtrsim 6\cdot 10^6$ for $T=T_c$ (left) or $n-m\gtrsim 10^7$ for $T=1$ (right)) 
is affected by non-linear effects. 
The variances show the behavior similar to the one already
discussed for the ferromagnetic Ising model.
A notable difference is that,
not only in the critical quench, but also in the sub-critical case, 
one has $\Delta ^{(\chi)}\simeq \Delta ^{(\chi)}_0$,
indicating that $\Delta ^{(\chi)}_r$ is always a sub-dominant
contribution. We have
explicitly checked that this happens also in quenches to lower
temperatures ($T=0.5, T=0.2$). 
This possibly indicates that the argument developed above
for ferromagnets cannot be straightforwardly extended
to the low temperature phase of spin-glasses and that
second order susceptibilities (or variances) may be used
as efficient tools to characterize this difference.

\subsection{Fredrickson-Andersen model} \label{fredand} 

As stated in ~\cite{lcz} and discussed further in \cite{noi_nonlin,noi_nonlin2} and 
in Sec.~\ref{model} of this Article,
the FDR of LCZ can be derived in full generality in the context of
Markovian dynamics, regardless of the model
Hamiltonian and of the choice of transition probability.
However, this claim has been questioned in~\cite{mayersollich} 
where the applicability of the LCZ algorithm
to the FA model was doubted.
However, the derivation of the LCZ relation as given above, or with different analytical
techniques in Refs.~\cite{noi_nonlin,noi_nonlin2,baiesi}, shows clearly
its general character where no particular assumptions on the specific system nor on the 
form of the perturbation are made. In order to illustrate 
this issue also numerically, we have carried out the computation the  
integrated response function with the LCZ algorithm
in the one-dimensional FA model.

The Hamiltonian of the FA model reads ${\cal H}(\sigma)=\sum_i\sigma_i$,
where the $\sigma_i$ are bimodal variables taking the values ($0,1$),
with $\sigma_i = 1$ for a mobile fluid region and $\sigma_i = 0$ an for an immobile one. 
Spins evolve according to transition  probabilities  obeying detailed balance, whose off-diagonal elements are
\be
w_i(\sigma'|\sigma)=[\epsilon(1-\sigma_i)+(1-\epsilon)\sigma_i]\lambda_i(\sigma),
\ee
where $\epsilon=1/(1+e^{1/T})$ is the equilibrium density and
$\lambda_i(\sigma)=\sigma_{i-1}+\sigma_{i+1}$ is a kinetic constraint that
preserves detailed balance, due to the independence from $\sigma_i$.
The integrated response function has been computed using both the SM and the LCZ algorithm.
In the SM, the effect of the external field amounts to replace $\epsilon$
with $\epsilon_i^h=1/(1+e^{(1-h_i)/T})$, whereas the quantity $B_i$ defined 
in Eq. (\ref{bi}) and entering
the LCZ relation is given by 
\be
B_i(\sigma)=[\epsilon(1-\sigma_i)+(1-\epsilon)\sigma_i]\lambda_i(\sigma)[1-2\sigma_i].
\ee

\begin{figure}
    \centering
    %\vbox to 8.5 cm {
    
   \rotatebox{0}{\resizebox{.45\textwidth}{!}{\includegraphics{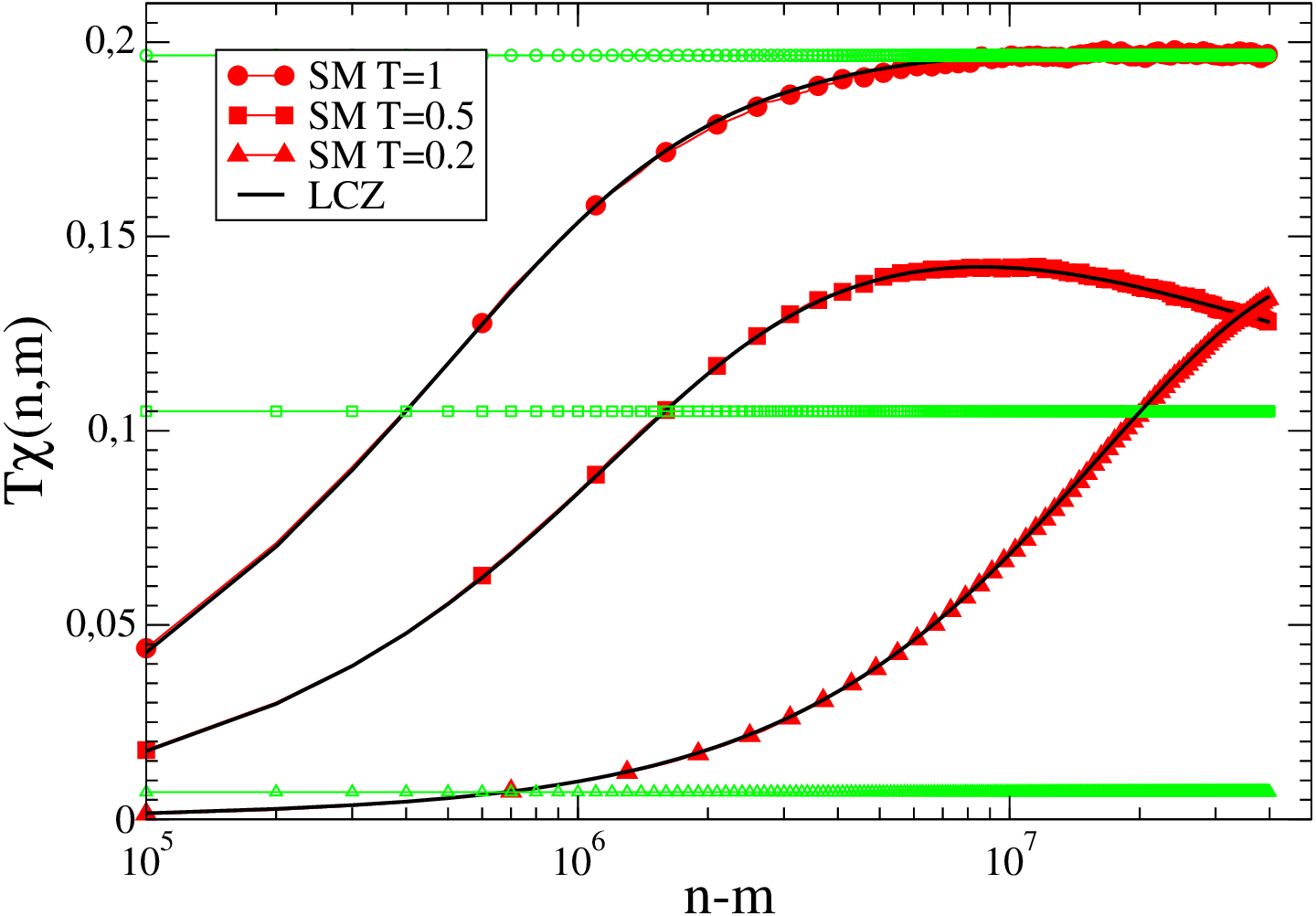}}}
    \caption{(Color online). Results of the numerical computation 
    of the integrated auto-response function 
    $\chi (n,m)$ for the one-dimensional FA model with $m=10^5$. The response functions computed
    with the SM (red filled symbols) and with the LCZ algorithm (continuous black lines),
    agree for the three temperatures. In the SM the external field is $h=0.01 T$,
    satisfying linearity. The equilibrium value (green empty symbols)
    is also plotted for each temperature.}
\label{fa}
\end{figure}

In Fig.~\ref{fa} the comparison between the data obtained with the two methods and for
three different temperatures, is presented.
The value of $h$ used with the SM 
was checked to be in the linear regime.
In particular we found that keeping the ratio $h/T=0.01$ constant 
satisfies the linearity requirement.
The agreement between the two computations is excellent in the whole time range
and for all the temperatures considered. Data for $T=1$ converge to
the equilibrium value whereas results for $T=0.5$ and $T=0.2$ exhibit
a non-monotonic behavior already observed in one-dimensional
kinetic constrained models \cite{Crisanti2}.
 At low temperatures the equilibrium value
is reached only asymptotically.  

\section{Conclusions and perspectives} \label{conclusions}

In this paper we have compared the FDR derived in a 
series of papers by CRT \cite{chatelain,rt} 
and LCZ \cite{lcz}. 
First, by re-deriving them in a unified
formalism we have pointed out that the distinction between
these FDR is due to a different choice of the 
perturbed transition probabilities. Actually,
even restricting to systems where detailed balance is
obeyed, for a given choice of unperturbed transition probabilities
an arbitrarity remains on the form of the perturbed ones,
which is parametrized by the function $m_j$ defined in Eq. (\ref{mpiccolo}).
In the case of Ising spins, selecting the value of $m_j$ corresponding to 
heat bath transition probabilities leads to the
CRT relation (\ref{ricci}). This FDR relates $R_{i,j}$
to a correlation function involving a $\delta$-function which
weights only a subset of the whole ensemble of unperturbed trajectories.
For this reason, the FDR of CRT, besides being limited to
Ising spins with heat bath transition probabilities, can only
be used in numerical simulations.  
On the other hand, making the choice $m_j=0$ of LCZ allows a 
further mathematical treatment leading to the FDR (\ref{LCZf}) 
where the response
function is related to standard unperturbed correlations. 
This makes this relation basically different
from those obtained in \cite{Berthier} (Eq. (\ref{berth})
and in other approaches (i.e. in \cite{Crisanti,Diez}), where the response
function cannot be expressed in terms of correlations of observables.
This makes the applicability of the LCZ relation in principle
not restricted to simulations. Furthermore, 
this FDR has a larger degree of applicability with
respect to the CRT, since it is not restricted to Ising spins nor to
heat bath unperturbed transition probabilities.
 
In the second part of this Article, 
we have studied the efficiency of the CRT and LCZ field-free methods. 
In order to do that, we evaluate analytically the
variances of the response function 
obtained with the SM or with the field-free methods.
It turns out
that, as far as the autoresponse function is considered, 
field-free methods are by far more efficient than
the SM. This combines with the advantage of having linearity
($h\to 0$) built in.
Moreover, we found that the LCZ algorithm is slightly more efficient (by a factor
$\sqrt 2$) than the method of CRT.  

We conclude by pointing out that the results contained in this paper
are not restricted to the framework of the efficient computation of
the response function. Indeed, we mention that the study of the variances
is closely connected to the issue of characterizing 
the fluctuation of two-time quantities in aging systems, a problem which has
received a good deal of attention recently \cite{c4}.
Moreover, as discussed at the end of Sec. \ref{variances}, 
the variance of the response function 
is also related to the second order susceptibility introduced
in \cite{huse,noi_nonlin,noi_nonlin2} for the study of cooperativity.

\section*{Acknowledgments}

F.Corberi, M.Zannetti and A.Sarracino acknowledge financial support
from PRIN 2007 JHLPEZ ({\it Statistical Physics of Strongly correlated
systems in Equilibrium and out of Equilibrium: Exact Results and 
Field Theory methods}).

\section{Appendix I}

We first prove Eq.(\ref{ssss}) 
\be
1-\langle \sigma_i(m+1)\sigma_i(m) \rangle =
\frac{1}{N}\left [1-\langle \sigma_i^W(m)\sigma_i(m) \rangle \right ].  
\label{a5}
\ee
To do that, we first observe that from Eq.(\ref{dsu})
\be
\langle 1- \sigma_i(m+1)\sigma_i(m) \rangle
=\frac{1}{N}\sum_{\sigma'',\sigma'} [1-\sigma_i''\sigma_i']
w_i\left (\sigma'' \vert 
\sigma' \right )P(\sigma',m).
\ee
Only the term $w_i\left (\sigma'' \vert 
\sigma' \right )$ in the transition probability between the time $m$ and
$m+1$ contributes. Hence the two configurations
$\sigma''$ and $\sigma'$ differ only for the spin on the $i-th$ site and
sum on $ \sigma''$
reduces to the sum on the two possible values $\pm \sigma_i'$.
Then, using the Heath bath form  for the transition probabilities,
\be
w_j\left (\sigma'' \vert 
\sigma' \right )
=\frac{1}{2} \left (1+\sigma_j'' \sigma_j^W(\sigma')
\right )
\label{HB2}
\ee
one obtains~Eq.(\ref{a5}). 

Next we prove Eq.(\ref{sssw}). Let us first proof that  
\be
\langle [\sigma_j(m+1)-\sigma^W_j(m)]\sigma^W_j(m)\delta_{I(m),j}\rangle=0.
\label{a1} 
\ee

From the definition one has
\be
\langle \left [ \sigma_j(m+1) -\sigma_j^W(m) \right ]
\sigma_i^W(m) \delta_{I(m),j}  \rangle = 
\sum_{\sigma'', \sigma'}
\left [ \sigma_j'' -\sigma_j^W(\sigma') \right ]
w_j\left (\sigma'' \vert 
\sigma' \right )\delta_{I(m),j}
 P(\sigma',m).
\ee
Because of the delta function, one has 
that the configurations $ \sigma'$ and $ \sigma''$ can
differ only for the spin in the $j$-th site. This implies  
that $\sigma_j^W(\sigma')=\sigma_j^W(\sigma'')$ and that the
sum on $ \sigma''$
reduces to the sum on the two possible values $\pm \sigma_j'$. 
The Heath-Bath form  (\ref{HB2}) for the $w$  then gives
\be
\sum_{\sigma_j''=\pm \sigma_j'}\left [\sigma_j''-\sigma_j^W(\sigma')
\right ] w_j\left (\sigma'' \vert 
\sigma'\right )=0,
\ee
which implies Eq.(\ref{a1}).
We next observe that from Eq.(\ref{pnm2}) 
\be
\langle \sigma_j^W(m)^2 \delta _{I(m),j} \rangle=
\frac{1}{N}\sum_{I(m)}\sum_{\sigma',\sigma}(\sigma_j^W)^2
w(\sigma' \vert \sigma)P(\sigma,m) \delta_{I(m),j}=
\frac{1}{N}\langle  \sigma_j^W(m)^2 \rangle .
\ee
Combining the above result with Eq.(\ref{a1}) one recovers Eq.(\ref{sssw}).

\section{Appendix II}

Here we prove that $R^{(2)}_{i,j}(n,m,m')=\langle r_{i,j}(n,m)r_{i,j}(n,m')\rangle$ is identically
zero for all the algorithms if $m\ne m'$.

\begin{enumerate}
\item{

{\it Standard method}

In the SM one applies a random magnetic field $h_i$, 
with $\overline h_i=0$ and $\overline {h_ih_j}=\delta_{ij}$,
and the response is obtained as
\be
R_{i,j}(n,m)=N\overline{\frac{\langle \sigma_i(n) h_j\delta_{I(m),j}\rangle}{h^2}}.
\ee
Therefore
\be
R^{(2)}_{i,j}(n,m,m')=\frac{N}{h^2} \delta_{m,m'}
\ee
}

\item{

{\it CRT relation}

In this case $r_{i,j}$ can be directly read off from Eq.~(\ref{flipnoflip}). 
From Eq.~(\ref{gj}) one has 
\be
\sum_{\sigma'',\sigma'}
w_i(\sigma'' \vert \sigma') f_i(\sigma'' , \sigma') 
[1-\delta_{\sigma'', \sigma'}]P(\sigma',m \vert \sigma,m')
=-\sum_{\sigma'} w_i(\sigma' \vert \sigma') g_i(\sigma') 
\delta_{\sigma'', \sigma'}P(\sigma',m\vert \sigma,m')
\ee
that gives
\be
\langle f_i(m)O(m')\rangle_{flip}+\langle g_i(m)O(m')\rangle_{no flip}=0,
\ee
for every generic observable $O$ computed at the shorter time $m'$.
Therefore, 
\be
R^{(2)}_{i,j}(n,m,m')=0 
\label{r202}
\ee
for $m'\le m-1$.}

\item{

{\it LCZ relation}

Reading $r_{i,j}$ from Eq. (\ref{LCZf}), using
the property~(\ref{b2}) one obtains that $\langle [\Delta\sigma_i(m)-B_i(m)]O(m')\rangle=0$,
for every generic observable $O$ computed at the shorter time $m'$.
Therefore one arrives again at Eq. (\ref{r202})
for $m' \le m-1$. }
\end{enumerate}

\section{Appendix III: Numerical impementation of alghorithms for the
  computation of integrated response functions}

\begin{enumerate}
\item{

{\it Standard method}

For each realization of the dynamical trajectory, 
a random magnetic field $h_i$ is assigned to each
site. $h_i$ is usually chosen from a bimodal distribution $h_i=\pm h$.
The evolution is then controlled by unperturbed transition probabilities until
the time $m$ and by the perturbed ones given in Eq.(\ref{dbh2}) for
later times. At the time $n$, the integrated response function is
computed according to
Eq.(\ref{eq57}). 

}

\item{

{\it CRT relation}

The integrated response function can be obtained from the space and
time integral of Eq. (\ref{ricci}) 
\be
\chi^{(CRT)}(n,m)=\frac{\beta}{N}\sum_i \langle \sigma_i(n) \Lambda_i(n,m),
\rangle
\label{AchiCRT}
\ee
with 
\be
\Lambda_i(n,m)=\sum_{l=m}^{n-1}\left [\sigma_i(l+1)-\sigma_i^W(l)
  \right ] .
\ee
In each realization of the dynamics, the quantity $\Lambda_i$ is
initially set to zero on each site. For all the timesteps $l\ge m$, 
$\Lambda_j$ is updated, via the relation
\be
\Lambda_j=\Lambda_j+\sigma_i(l+1)-\sigma_i^W(l), 
\ee
only on the site
$j=I(l)$ where the flip of the spin has been attempted.   
Elsewhere $\Lambda_i$ is left unchanged. 
At each time $n\ge m$, $\chi(n,m)$ is then computed according to Eq.(\ref{AchiCRT}).
}

\item{
{\it LCZ relation}

From Eq. (\ref{LCZf}) one immediately obtains
\be
\chi^{(LCZ)}(n,m)= \frac{\beta}{2N}
\sum_i \left [ \langle \sigma_i(n) \sigma_i(n) \rangle -
\langle \sigma_i(n) \sigma_i(m) \rangle +
\langle  \sigma_i(n-1)A_i(n,m) \rangle \right ]
\label{AchiLCZ}
\ee
with 
\be
A_i(n,m)=\frac{1}{N}\sum_{l=m}^{n-1}B_i(l) .
\ee
The quantity $\langle \sigma_i(n)\sigma_i(m)\rangle $ is the usual two
time correlations function. Concerning the evaluation of $A_i$, the
basic observation is that, according to Eq.(\ref{eq68}), $B_i$ only
depends on the spin $\sigma_i$ and on the spins interacting with it. In
particular, for the models considered in this paper, $B_i$ depends on
$\sigma _i$ and on its nearest-neighboring spins.   
The evaluation of $A_i$ proceeds as follows. At the time $m$, $B_i$ is
evaluated on each site according to Eq.(\ref{eq68}), $A_i$ is set to zero and $l_i$ is set to
$m$. $l_i$ represents the time where the last evaluation
of $B_i$ has been performed. $B_i$ is then left unchanged on all sites
until a spin flip occurs. If the spin flip occurs on site $I(l)$ at time $l$,
$A_j$ is updated as
\be
A_{I(l)}=A_{I(l)}+B_{I(l)}(l_j) \left(l-l_j\right)/N.
\ee   
$l_j$ is then set to $l$ and the new value of $B_j$ is evaluated.     
The same procedure is repeated for all the spins interacting with $\sigma _{I(l)}$,
leaving unchanged $A_i$, $B_i$ elsewhere.
In the end $A_i(n,m)$ is obtained via the relation
\be
A_i(n,m)=A_i+B_i(l_i) \left(n-l_i\right)/N,
\ee
and the integrated response function is computed  according to
Eq.(\ref{AchiLCZ}).
}

\end{enumerate}

\end{document}